# Strategic API Analysis and Planning: APIS Technical Report

## Authorship

**Research Authors**
Jennifer Horkoff, jenho@chalmers.se
Juho Lindman, juho.lindman@ait.gu.se
Imed Hammouda, imed.hammouda@cse.gu.se
Eric Knauss, eric.knauss@cse.gu.se

**Industrial Authors**
**We thank the representatives from the six companies who provided example cases and made major contributions to the framework.**

## How to Read and Use this Document

This document reports findings from the APIS (Strategic API Analysis and Planning) project. The overall findings are collected as part of a framework of individual topics with their own methods. After an overview, each topic is presented, along with a description of the intended benefits of that particular method, and some guidance on when it may be appropriate to use or not to use the method.

Although the reader is welcome to read the document as a whole, not every topic or method may be relevant at a particular time. Therefore, this document is meant to act as a reference, allowing for the application of particular methods when applicable.

# Table of Contents

# 1 Motivation: Strategic API Analysis and Planning

Traditionally, software APIs (application programming interfaces) have been viewed from a technical perspective, as a means to separate implementation from functional calls, and as way to define a contract of software functionality. The technical benefits of APIs have been reported in numerous studies [1]. APIs provide a useful interface for service provision, customer access, and third-party development. The use of APIs provides not only control but also stability to encourage use and reduce abuse. In addition to solving technical challenges related to API management, APIs are considered more of an artifact that must be continuously enhanced [2]. Several reports from the industry offer useful practical design considerations for APIs, including advice on collecting usage data, monetization strategies, and at what point to open an API to external parties [3, 4, 5, 6]. Although this advice can be useful, the focus is still often on the API, only without considering the role the API plays in the wider organization or how it fits into an organizational strategy. Our industrial experiences show that more information is needed about the challenges and best practices of API design and management in an organizational context. Furthermore, it has become apparent that APIs are able to play a key role as part of a strategic business plan for software-intensive companies, noted by both academia [1, 2], and industry [4-6].

In this work we report our work with companies to build a framework that synthesizes and summarizes API strategies from several perspectives, including:
- Strategic API Lifecycle
- Lifecycle Stage Characteristics
- Lifecycle Use Cases and Transition Points
- API Layered Architecture
- BAPO
- Value Modeling
- Layering Value Models
- Ecosystem Mapping with Goal Models
- Layering Goal Models
- API Metrics Framework
- Goal Models with Metrics
- API Governance Guidance

Our research was carried out within an industry–academia collaboration (see Methodology in Section 8.6). The framework was developed iteratively through workshops and discussions with company partners. We show how conceptual frameworks can be used to drive API strategy development, drawing on our research work in organizations with several companies in practice. In this document, we aim to present our results in a usable format.

## 1.1 Further Reading

[1] C. R. B. de Souza and D. F. Redmiles, "On the roles of apis in the coordination of collaborative software development," Computer Supported Cooperative Work (CSCW), vol. 18, no. 5, p. 445, 2009.

[2] I. Hammouda, E. Knauss, and L. Costantini, "Continuous api-design for software ecosystems," in Proc. of 2nd Int. WS on Rapid and Continuous Software Engeering (RCoSE '15 @ ICSE), Florenz, Italy, 2015.

[3] Apigee, "Is your API naked? 10 roadmap considerations for API product and engineering managers," 2010. [Online]. Available: https://apigee.com

[4] Nordic API, "Developing the API mindset: A guide to using private, partner, & public APIs," 2015. [Online]. Available: https://nordicapis.com
[5] IBM Institute for Business Value, "Evolution of the API economy. adopting new business models to drive future innovation," 2016. [Online]. Available: https://www.ibm.com/common/ssi/cgibin/ssialias?htmlfid=GBE03759USEN
[6] Oracle Communications, "Making money through API exposure. enabling new business models," 2014. [Online]. Available: http://www.oracle.com/us/industries/communications/comm-makingmoney-wp-1696335.pdf

## 2   APIS Use Cases

In this section, we list the possible use cases for the various methods of the APIS framework and match them to potential sub-sections of the report.  The idea is to give early recommendations on which particular sub-methods can be helpful, depending on the needs and use cases of the organization.   Note that the list of possible use cases is not necessarily complete.

*Table 1   APIS Use Cases Mapped to Specific APIS Framework Methods*

| Use Case | Method |
| --- | --- |
| Identifying API roles, users, assets | Layered architecture (Section 5.1) |
| Identifying API boundary objects | Layered architecture (Section 5.1) |
| Identify business motivations for API | BAPO (Section 5.3), Goal Modeling (Section 7.2), Value modeling (Section 7.1) |
| Identify the API from an architecture perspective | BAPO (Section 5.3) |
| Identify the API from a process perspective | BAPO (Section 5.3) |
| Understand how API contributes to or shapes organization | BAPO (Section 5.3) |
| Identify lifecycle stage of an API | Lifecycle characteristics (Section 6.1) , Lifecycle model (Section 6.2) |
| Compare and contrast API lifecycles | Lifecycle characteristics (Section 6.1) , Lifecycle model (Section 6.2) |
| Look for inconsistency in API behavior | Lifecycle characteristics (Section 6.1) , Lifecycle model (Section 6.2) |
| Promote awareness of lifecycle transition points | Lifecycle transition points (Section 6.3) |
| Improve lifecycle transition points for an API | Lifecycle transition points (Section 6.3) |
| Determine governance for lifecycle transition points for an API | Lifecycle transition points (Section 6.3) |
| Understand API ecosystem | Goal Modeling (Section 7.2), Value modeling (Section 7.1) |
| Support ecosystem scoping | Goal Modeling (Section 7.2), Value modeling (Section 7.1) |
| Identify value of an API for different actors | Value modeling (Section 7.1) |
| Find problematic or missing value flows | Value modeling (Section 7.1) |
| Understand governance flows, actions | Goal Modeling (Section 7.2), Value modeling (Section 7.1) |
| Find missing actors and resources in ecosystem | Layering goal models (Section 7.3) |
| Understand role of actors in an ecosystem from an API perspective | Layering goal models (Section 7.3) |
| Understanding how roles of each actor changes in an ecosystem depending on API of focus | Layering goal models (Section 7.3) |
| Understand the desired qualities of the API ecosystem | Goal Modeling (Section 7.2) |
| Understanding the goals and dependencies of an API ecosystem | Goal Modeling (Section 7.2) |
| Designing an API ecosystem | Goal Modeling (Section 7.2), Value modeling (Section 7.1) |
| Evaluating options for how to participate in an existing API ecosystem | Goal Modeling (Section 7.2) |
| Evaluating the health of an API ecosystem | Goal Modeling (Section 7.2) |
| Finding relevant API metrics | Determining API metrics (Section 8.2) |
| Better understand role and impact of API metrics | Layered metrics (Section 8.4) |
| Find missing metrics | Layered metrics (Section 8.4) |
| Identifying API design-specific metrics | Automating design metrics (Section 8.5) |
| Considering automation of API design-specific metrics | Automating design metrics (Section 8.5) |
| Understanding why metrics are being measured | Linking Metrics to Goal Models (Section 8.3) |
| Understanding who receives metric information and why | Linking Metrics to Goal Models (Section 8.3) |
| Evaluating satisfaction of goals via API metric values | Linking Metrics to Goal Models (Section 8.3) |
| Understanding governance roles and process | API Governance guidance (Section 8.1) |
| Identifying weaknesses and problems in API governance | API Governance guidance (Section 8.1) |
| Identifying consideration and components of an API governance process | API Governance guidance (Section 8.1) |
| Determining which parts of an API to govern | API Governance guidance (Section 8.1) |
| Determine the level of governance needed for an API or parts of an API | API Governance guidance (Section 8.1) |

# 3 Running Examples

We have developed the framework along with five industrial partners providing example API cases. We describe briefly the companies and their API cases in this section. We will use the company examples to illustrate methods and topics throughout the document.

## 3.1 Company Participants

### 3.1.1 Company A

Company A is global firm operating in the area of network video cameras, currently providing network video products which are installed in public spaces (for example train stations and universities) and business areas (for example casinos and retail stores). The company adds value to their cameras by providing an API to enable their customers create their own applications where they use the generated camera data. Planned provision of a Cloud API will serve the needs of customers, increase customer satisfaction and customer retention.

### 3.1.2 Company B
*B* is a large telecommunications company with more than 110,000 employees. Company offer services, software and infrastructure in information and communications technology (ICT) for telecommunication and networking equipment.

### 3.1.3 Company C
*C* is an international company developing different mechatronic devices for its global customers around the world and is moving towards the implementation of a new suggested API ecosystem workflow, is expected to solve the bottlenecks in their current business process. The company wants to be able to analyze the effects these changes have on their API ecosystem design.

### 3.1.4 Company D
*D* provides primarily packaging for liquid and food products but also a range of processing and packaging technologies in a broader array of products. Company supplies databases that contain information their customers use to generate reports about manufacturing tasks, e.g., quality control reports. In order to make the generation of these reports easier, the organization aims provide an API to access the needed data easily and generate reports much faster.

### 3.1.5 Company E
*E* is an international company involved in many sectors, including consumer goods.  We worked with the company's software development center, where they focus on mobility services for devices.

### 3.1.6 Company F
*F* is a large, international technology company working in several areas including manufacturing, hardware and software.   In their case, the APIs of focus are rather large, and have and API Management processes established. They focus on further improvement e.g. in metrics, or governance.

## 3.2 API Examples from Companies

### 3.2.1 Cloud API
Value is created by providing a cloud API to enable third-party developers to create their own applications using application data. One API of focus, a cloud API, was in the planning stages during most of the development of the framework. Key issues included the role of 3rd party developers in the external cloud API.

### 3.2.2 Network API

A network API for configuring and operating Video devices.

### 3.2.3 Design Rules API
The study was conducted in cooperation with an internal unit that was working on developing global software design rules handling to state, fault, and alarms.  The API of focus was in use.

### 3.2.4 Device API

A new API facilitating access to devices via mobile and other cloud devices. The API would be generally for internal use. Key issues include supporting differing development speeds of different components.

### 3.2.5 Platform API

This case involves a large framework or platform which was developed internally, but gradually migrated to more off-the-shelf-components. This project was interested in metrics to help with refactoring decisions.

### 3.2.6 Framework APIs

Desktop APIs are part of a large framework where effort has been spent automating and controlling the API processes and development, including introducing API-specific internal training.

### 3.2.7 Product API

The investigated API is mainly internal and related to the reuse of common function signatures across products. The API of focus was in partial operation. Key issues included governance, and the role of the API owner.

### 3.2.8 Reporting API

Reporting API contains information customers use to generate reports about manufacturing tasks (e.g., quality control reports). In order to make the generation of these reports easier, the organization aims to provide an API to access the needed data easily and generate reports much faster. The API of focus, a reporting API, was in the planning stages.

### 3.2.9 REST APIs

This part focused on roughly 50 internal and 20 external, which aims for similar governance and metrics throughout.

### 3.2.10 Service API

Internal set of scripts automating common tasks involving company hardware. The API is internal, but geographically distributed. Part of the API of focus had been developed, while the part focusing on automation was under development. Key issues include avoiding risks by limiting direct access to hardware while still providing needed functionality.

### 3.2.11 Technology APIs

This example focuses on a set of related APIs related to specific technologies. The APIs are used internally and range from very active to near retirement. Key issues included the retirement of API functionality and governance of API change.

## 4 APIS Framework Overview

In this section, we will provide an overview and summary of each topic and method as part of the framework. This will allow the reader to pick and choose which sections are applicable to their case. Further details are provided within the specific subsections. An overview of the API Framework is shown in Figure 1.

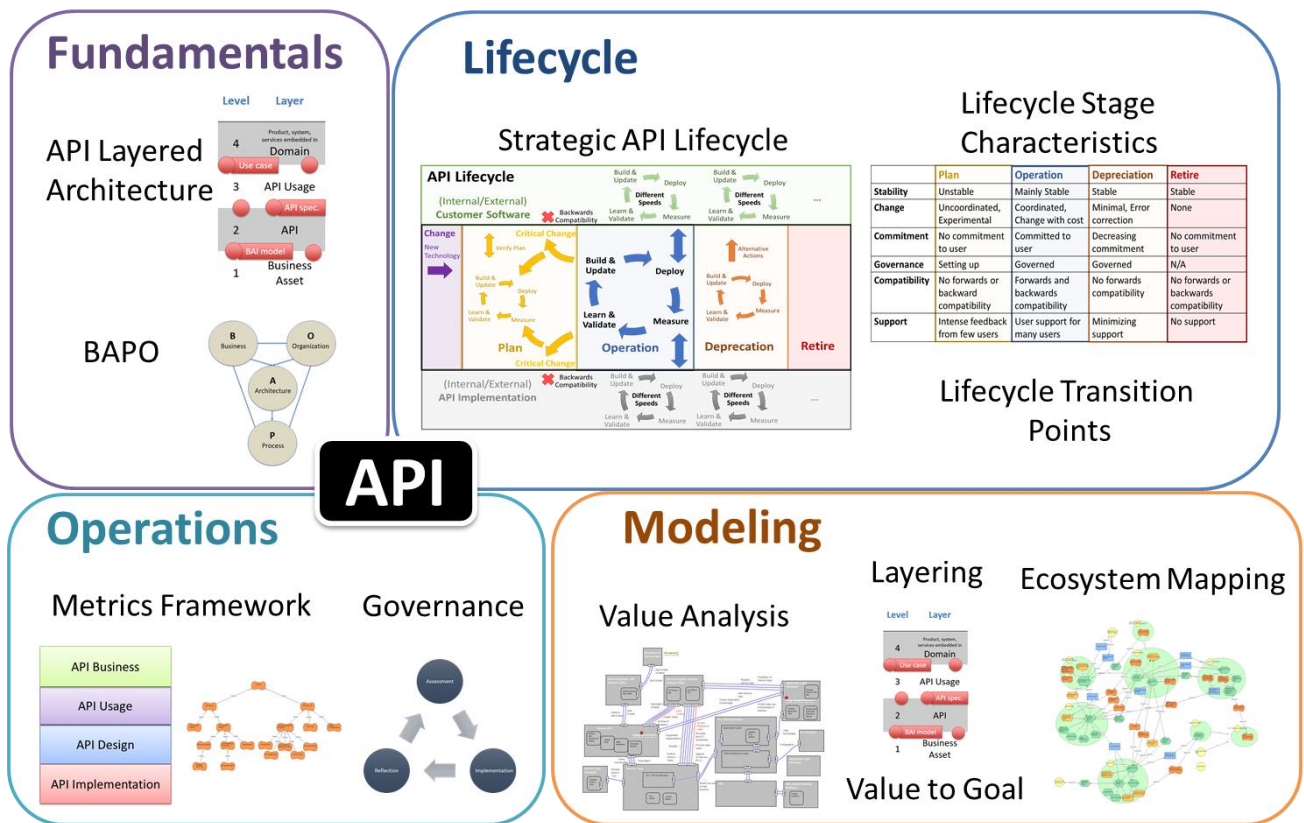

*Figure 1: Overview of API Framework showing Framework Topics*

## 4.1 Fundamentals

### 4.1.1 API Layered Architecture

Layered view are common in software/hardware architecture. Here we take a more strategic and organizational view of API layers, considering the business, software usage, API, and asset layers.

A consideration of an API Layered Architecture provides the following ***benefits***:
- Classifying roles, groups, organizations and technical elements as part of the business motivation, users, API or asset.
- The above allows one to explicate API users, the assets that APIs protect, and those to seek benefits from the API.
- The classification often reveals gaps when certain types of roles have not been thoroughly accounting for, helping to increase the completeness of the view of the API Ecosystem.

More detail can be found in Section 5.1.

### 4.1.2 BAPO

We view APIs from the lens of the BAPO (business, architecture, process, and organization) framework. This viewpoint provides the following ***benefits***:
- Clearly identify the business motivations, actors, and users involved in the API.
- View the API from an architecture perspective, considering alignment with software structures.

- View the API from a process perspective, to understand what role the API plays in existing processes.
- Understand how the API contributes to organizational aspects.

More detail can be found in Section 5.3.

## 4.2 API Lifecycle Analysis
### 4.2.1 Lifecycle Stage Characteristics
We present the characteristics of APIs in each stage of the cycle in terms of stability, freedom, support, etc. **Benefits** include:
- Identify the lifecycle stage of an API to aid in strategic planning and potential transitions
- Comparison across different APIs
- Look for inconsistency in API behavior, e.g., if an APIs is unstable but has a high level of support

More information can be found in Section 6.1.

### 4.2.2 Strategic API Lifecycle
We have worked to understand the typical lifecycle stages through which an API progresses, including planning, operation, deprecation and retirement. Our lifecycle model covers the view of the API, API user and API implementer, considering the role of versioning, new technologies and other forms of change. Our lifecycle is strategic as we consider business and organizational aspects when capturing API phases.

A consideration of API lifecycle provides the following *benefits*:
- Allows API analysts and stakeholders to determine at which part of the lifecycle an API of interest lays. This helps to make strategic decisions about lifecycle decisions (e.g., is it time to go operational? Depreciate an API?)
- This also helps to have a common vocabulary in which to discuss the status of an API.
- The lifecycle framework also provides a way to compare and contrast the relative development of multiple APIs.

More information on our Strategic API Lifecycle framework can be found in Section 6.

### 4.2.3 Lifecycle Use Cases and Transition Points
This idea with this topic is to understand particular use cases for APIs in each stage of lifecycle development, including triggers or transition points which indicate progression to the next stage. **Benefits** include:
- Promote awareness of lifecycle stage transition points
- Improving transition points between lifecycle stages
- Consider governance of transition points

More information on our Strategic API Lifecycle framework can be found in Section 6.3.

## 4.3 API Analysis with Modeling
### 4.3.1 Value Modeling

API ecosystems can be considered in terms of value flows, focusing on capturing the relevant ecosystem actors, and how they provide value to each other.   This technique makes use of an established modeling technique, e3 value modeling, describing how to apply this technique to the specific case of API ecosystems.  **Benefits** of the method include:
- A visual representation of the API ecosystem
- Supporting ecosystem scoping, understanding which actors are relevant from an API perspective
- Identifying value provide to and from the API and related actors
- Identifying non-reciprocated and problematic value flow, helping to find issues and problems

More detail can be found in Section 7.1.

### 4.3.2   Ecosystem Mapping with Goal Models

Just as one can map the ecosystem using tables (Section 4.1.1, 5.1) or value models (Section 4.3, 7.1), a goal-based perspective can also be used, focusing on why each interaction or API-related task is needed.  This also provides a view of the API ecosystem actors, but instead of capturing value flows, it captures dependencies between actors linked to the satisfaction of internal goals and qualities.  Goal modeling also captures task decomposition and strategic alternatives.  The **benefits** of this method include:
- An alternative visual view of an API ecosystem
- Articulating the business and individual goals of each actor in the ecosystem
- Capturing goal hierarchies
- Capturing and evaluating alternative strategies
- Linking interactions (dependencies) between actors to actor goals, capturing why
- A focus on API qualities or NFRs in an ecosystem

More detail can be found in Section 7.2.

### 4.3.3   Layering Goal Models

One can get further value out of API ecosystem goal models by sorting the actors into the API layers from Section 4.1 and 5.   This allows one to view the ecosystem in terms of API layers, finding missing actors and considering changing perspectives depending on the API of focus.

More detail is found in Section 7.3.

### 4.3.4   Transforming Value Models to Goal Models

One can use API ecosystem modeling with goal models by drawing the models from scratch, or one can start by transforming the value model into an incomplete version of a goal model.   This section describes how to do the latter.   The benefits including having the two views of the API ecosystem (value and gal model) consistent, and simplifying the process of creating goal models.

More detail is found in Section 7.4.

## 4.4   Operations

### 4.4.1 API Governance Guidance

API governance is one of the most important mechanisms to ensure performing and compatible API management in an organization. Several aspects of governance were discussed in the interactive collaborations and during the modelling sessions with the companies. We have gathered multiple examples of different governance practices, and compare these using visual value models. We discussion clear governance practices that suit the different API layers and stages of life-cycle.

The ***benefits*** include, but are not limited to:
- Understanding the main components of functioning multi-level API governance in organizational setting
- Identifying the weaknesses of the current governance mechanisms related to API
- Strengthening these mechanisms
- Help in formalizing the transition points of API governance (or choosing not to formalize)

More details and examples can be found in Section 8.1.

### 4.4.2 Determining API Metrics

Measuring the operations of an API from a strategic and ecosystem perspective can provide insight into the health of the API. ***Benefits*** of this method include:
- Examples of API metrics used in practice
- Understanding the challenges specific to API measurements

More detail can be found in Section 8.2.

### 4.4.3 Linking Metrics to Ecosystem Models

The metric classification and exploration from Sections 4.4 and 8 can be mapped onto the goal models created with the methods in Sections 4.3.2 and 7.2. Existing work has added indicators (e.g., Key Performance Indicators (KPIs) to goal models, working out how they fit with the existing concepts. This allows one to map API metrics to goals and tasks collected in the goal model, answering "why" for each metric, and understanding how the metrics and their associated goals fit into the larger ecosystem. ***Benefits*** of this method include:
- Working out the rational or goal for each metric
- Assigning metrics to interested ecosystem actors
- Organizing metrics into a hierarchy
- Possibility to support qualitative and quantitative reasoning over goals using metric data

More detail can be found in Section 8.3.

### 4.4.4 Layered Metrics

We have made progress in identifying different classes of API measures as per our layered architectures. ***Benefits*** of this method include:
- Classifying API metrics into the different API layers for better understanding of metrics
- Helping to find potentially missing metrics
- Understanding the impact of metrics
- Provide a checklist of potential API metrics at each level

More detail can be found in Section 8.4.

### 4.4.5 Automating Design Metrics
API design metrics are both particularly important and difficult to automate. We list design-specific methods and consider their potential for automation. ***Benefits*** of this method include:
- Considering metrics specific to the design layer
- Consideration of potential level of automation of API metrics

More detail can be found in Section 8.5.

# 5 Fundamentals
## 5.1 API Layered Architecture

### 5.1.1 Intended Benefits
- Classifying roles, groups, organizations and technical elements as part of the business motivation, users, API or asset.
- The above allows one to explicate API users, the assets that APIs protect, and those to seek benefits from the API.
- The classification often reveals gaps when certain types of roles have not been thoroughly accounting for, helping to increase the completeness of the view of the API Ecosystem.
- The general understanding of the API and its ecosystem can be explicated and shared for discussion

### 5.1.2 When to use this method
When the overall view of the API is unclear. This method can help to get an early, high-level view of the API ecosystem.

### 5.1.3 When not to use this method
When the API, its users, usage scenarios and assets are generally known and well understood by all.

### 5.1.4 Description
Business needs are increasingly a driving force in the development of Application Programming Interfaces (APIs) that allow downstream developers to access (business) assets, for example, data and services. Developers may use an API internally, or APIs can be open to external third parties. API providers expose assets to internal or external parties in order to allow them to develop value-adding applications, where the API controls access to business assets.

To consider these aspects in a systematic way, we have created a layered view of APIs. Layered view are common in software/hardware architecture. Here we take a more strategic and organizational view of API layers, considering the business, software usage, API, and asset layers. The layered API model is shown in Figure 2. The model consists of four layers with boundary objects dealing with the interfaces between layers. The layers can be used together with or without considering the boundary objects explicitly. We describe each layer and corresponding boundary objects in the following.

### 5.1.4.1 Domain

The domain layer covers the business and human aspects of the API usage. This includes the overall product the API may be supporting, related services, users, using organizations, and organizational aspects of the company producing the API. Between the Domain and the API Usage layer lies the boundary object of Use cases, describing how the API will be used in practice.

### 5.1.4.2 Use Cases

It is useful to explicitly list the use cases in the domain which motivate the software using the API. These are the domain-facing use cases which necessitate the use of the API by the usage software.

### 5.1.4.3 API Usage SW

The API Usage layer focuses on the Software or technology which directly uses the API.

### 5.1.4.4 (Optional) API Specification (Spec.)

This is a list of the functionality provided by the API. It can be very similar to the use cases, as the functions should facilitate the use cases, thus this list is optional. This is the signature or specification of API function calls, including how to use the API.

### 5.1.4.5 API

The API layer represents the API itself, including the API implementation. Between the API and Business Assets is the BAI (Business Asset Interface) model, a high-level abstraction of the business assets, as in a database schema or metamodel.

### 5.1.4.6 Asset Model

The API protects a number of business assets, but not all assets or all attributes or information associated with these assets may be exposed. The asset model describes what part of the assets are exposed through the API.

### 5.1.4.7 Business Asset

APIs are meant to protect and strategically expose some asset. The asset can be computational, informational or cyber-physical.

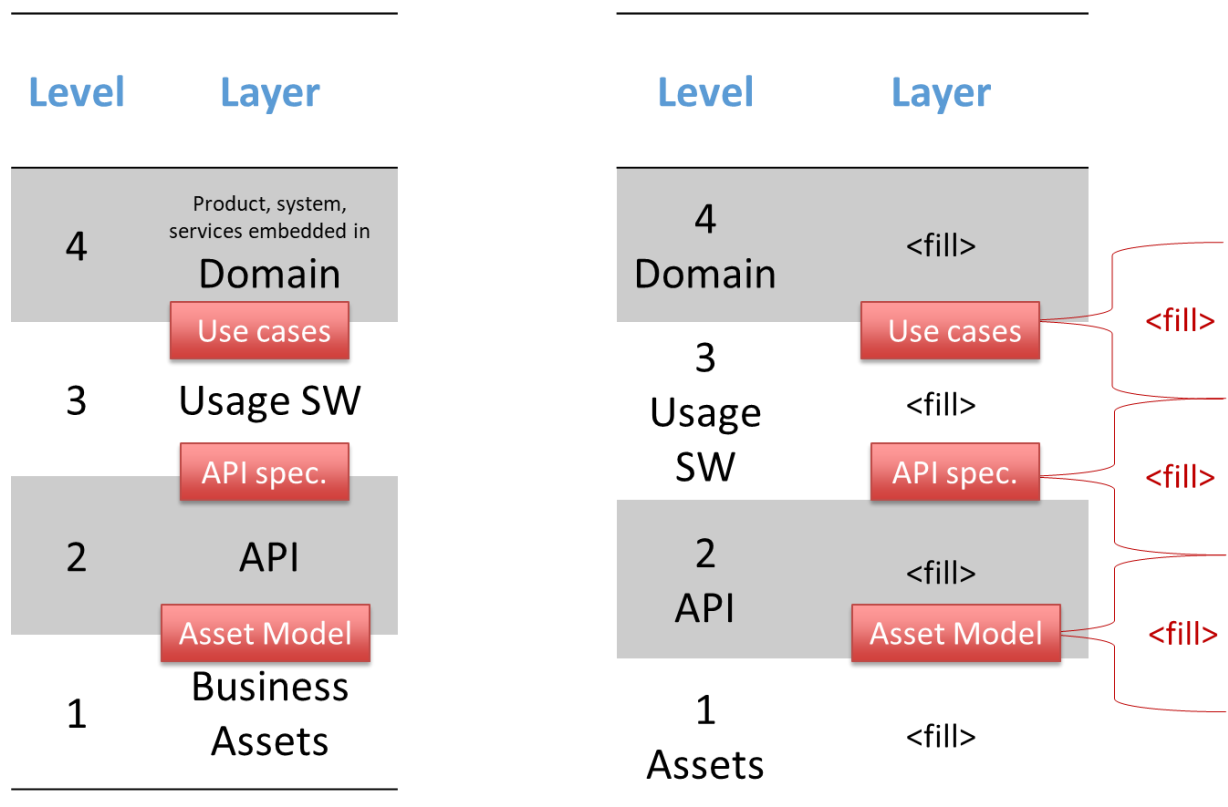

*Figure 2: API Layered Architecture (summary on left, fillable version on right)*

### 5.1.5 How to Apply

The basic idea is to map the ecosystem of the API to the various layers. For each API of interest, one should be able to list the technical or organizational contents of each layer. The following guiding questions can help:

- Who are the users of the API, i.e. who uses the software that uses the API? What organization do they belong to? In which products or services is the API used?
    - What are the use cases of the API in practice?
    - What is exchanged in the boundary between the domain the API software usage? What sort of descriptions or commitments?
- What are the use cases of the software using the API?
- What software or technology uses the API?
    - What information is exchanged between the API and the usage software? What agreement? Contract?
    - Is the software using the API internal or external?
- What functionality does the API provide?
    - Where is this functionality explained?
- Where is the implementation of the API? Where is it stored?
    - Is the software implementing the API internal or external
    - What agreement or information is passed from the API implementer to the API?
- Which parts of the assets are exposed via the API? Is there an agreement here?
- What business assets does the API protect?

### 5.1.6 Examples

We show the mapping of various aspects of some of our example APIs to the API layers as described in Table 2.

| Level | Layer | Device API | Cloud API | Product API | Technology API | Service API |
|---|---|---|---|---|---|---|
| 4 | Domain | Embedded device, cloud, phone, app users | Many, anyone using enhanced camera data | Device Communication, Product Development | Product functionality | Service Engineers, Developers |
| 3 | API Usage | Cloud service, smart phone, apps | Cloud Services | Internal consumer, Mobile App | New and revised Features | Service Tools, Service Apps |
| 2 | API | Embedded Device API | Cloud API | Device protocol/ Profile Spec | Function Module API | Application Layer API |
| 1 | Business Asset | Embedded Device | Camera Data | Pump Data | Function Modules | Machine, machine functions |

*Table 2: Mapping of API Examples to API Layers*

### 5.1.7 Summary
A layered architecture involving both organizational and technical aspects can help to provide a high-level overview of the API, filling in knowledge gaps, and developing a shared understanding of the API.

### 5.1.8 Further Reading
Y. Yoo, O. Henfridson, and K. Lyytinen, "The new organizing logic of digital innovation: An agenda for information systems research," Information Systems Research, vol. 21, no. 4, pp. 724–735, 2010.

## 5.2 API Openness

### 5.2.1 Intended Benefits
- Determine and classify the level of openness of an API
- Understand the finite nature of API-related resources

### 5.2.2 When to use this method
When the overall view of the API is unclear. This method can help to get an early determination of access to the API and the nature of the API resources.

### 5.2.3 When not to use this method
When the API and it's level of openness are well understood, or when the resources underlying or protected by an API are infinite.

### 5.2.4 Description
As an initial exercise, the organization should consider the degree of openness of the API and the API ecosystem.  For this, we have found it useful to refer to Ostrom's framework describing common pool resources, shown in Figure 3.   Here exclusion considers how exclusive access to the resource will be, while subtractability considers how likely the resource is likely to deplete.  The framework considers mainly physical resources, thus the notion of subtractability may not apply to something like an API, although other technical resources like server bandwidth are finite and subtractable.

Using this framework, an organization should answers:  is the API totally private?   Or is it a "club" situation where certain key partners can use and perhaps modify and API?  Or is the API open?  Are the resources protected by an API or the computational resources of the API itself finite?  Likely to deplete?

|  | Subtractability | |
|---|---|---|
|  | Low | High |
| **Exclusion: Difficult** | **Public Goods** Sunset Common knowledge | **Common-Pool Resources** Irrigation systems Libraries |
| **Exclusion: Easy** | **Roll or Club Goods** Day-care centers Country clubs | **Private Goods** Doughnuts Personal computers |

*Figure 3 Ostrom's Framework on Common Pool Resources*

### 5.2.5 Further Reading
1. Ostrom, Elinor. *Governing the commons*. Cambridge university press, 2015.

## 5.3 BAPO

### 5.3.1 Intended Benefits
- Consider an API from multiple perspectives
- Helps to facilitate an initial understanding of an API from a broader view, beyond technical

### 5.3.2 When to use this method
- When the organization has not previously considered the API from a non-technical perspective
- When one wants to map out and agree upon aspects like business strategy, architecture, process and organization

### 5.3.3 When not to use this method
- When the strategy and bigger picture of the API is relatively well-known.

### 5.3.4 Description (from [3])

All of our company respondents have emphasized that API development should not be an accidental or ad-hoc activity, but rather a systematic process supported by an integrated environment of policies, methods, tools, resources, etc. America et al. [2] argue that for a development method to achieve the best possible fit, four interdependent software development concerns shall be considered: Business, Architecture, Process, and Organization (often referred to as the BAPO model). Such a model helps to consider a wide range of strategic concerns in API strategy design.

From the lens of BAPO, one could identify the following kinds of challenges related to API design and development. We illustrate the discussion with example questions raised by our industrial partners.

Business
This perspective addresses concerns of making APIs a business capability and a business development model. Typical questions include: How API design drives business level decisions such as vendor selection and how to generate business value out of API usage?

Architecture
The main concern of this perspective is to investigate the technical issues associated with API design and development. Typical Questions include: How to manage API versioning (e.g. side-by-side deployment of different versions)? How to design APIs for extension? How to check backward compatibility of APIs between different versions?

Process
This perspective raises concerns related to identifying roles, responsibilities and relationships in order to respond and act more effectively within the API team. Typical Questions include: How to build a governance strategy for APIs that takes into consideration design phase issues (e.g. interface definition, quality trade-offs, reuse, documentation)?

Organization
This perspective covers concerns related the organizational aspects of API strategy, i.e. mapping the identified roles and responsibilities to existing organizational structures. Typical questions include: How to achieve better alignment between interacting organizational architecture units? How to build an effective API team? How to adapt existing organizational frameworks to the API context?

### 5.3.5 Examples

In early rounds, we have applied BAPO along with the layered architecture from Section 5.1 and the common goods model from Section 5.2 to map out example partner APIs from a high-level. In this way, we answered the following questions:
- How open is the API?
- At each layer, what are the issues or challenges of the API?
- Are these issues related to Business, Architecture, Process, or Organizational aspects?

We present examples for some of our partner organizations in Figure 4, Figure 5, Figure 6, and Figure 7.

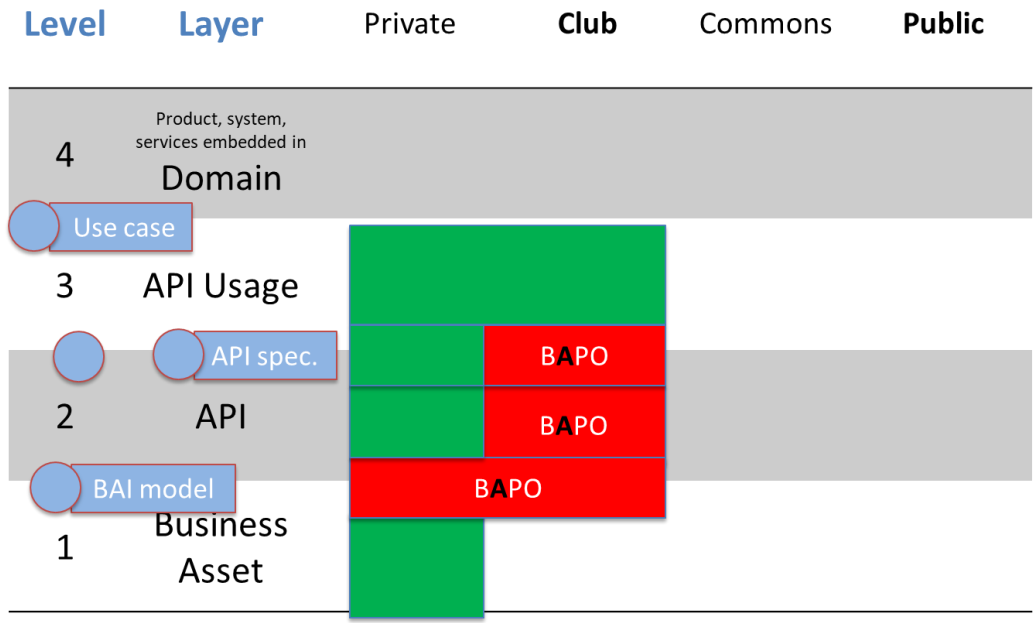

*Figure 4: Example of APIS layers vs. Common Pool vs. BAPO analysis for Design Rules API*

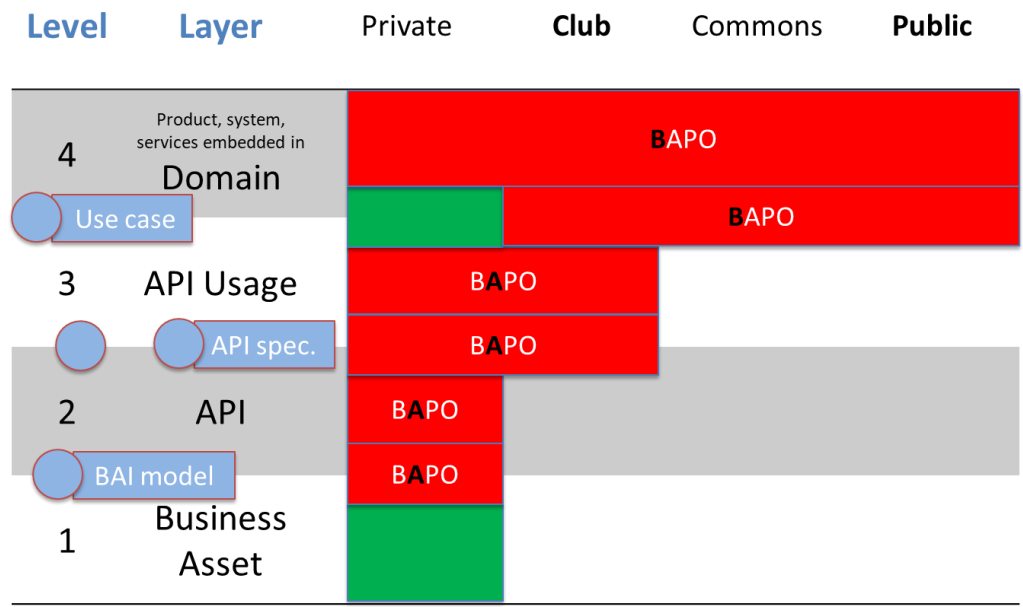

*Figure 5: Example of APIS layers vs. Common Pool vs. BAPO analysis for Cloud API*

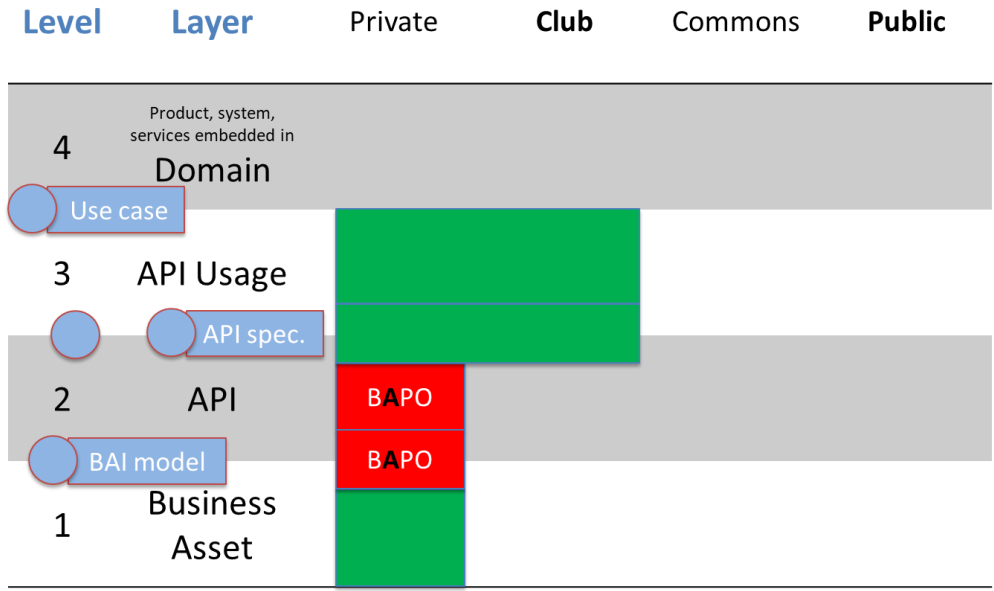

*Figure 6: Example of APIS layers vs. Common Pool vs. BAPO analysis for Reporting API*

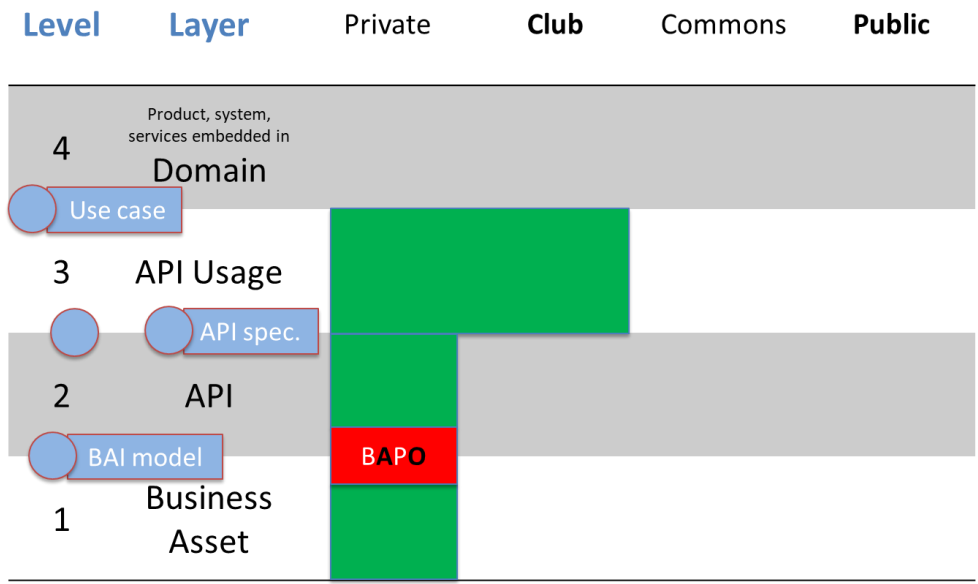

*Figure 7: Example of APIS layers vs. Common Pool vs. BAPO analysis for Product API*

### 5.3.6 Further Reading
1. Van Der Linden, Frank, et al. "Software product family evaluation." International Conference on Software Product Lines. Springer, Berlin, Heidelberg, 2004.
2. P. America, H. Obbink, R. van Ommering, and F. van der Linden, CoPAM: A Component-Oriented Platform Architecting Method Family for Product Family Engineering. Boston, MA: Springer US, 2000, pp. 167–180.
3. Lindman, J., Horkoff, J., Hammouda, I., & Knauss, E. (2018). Emerging Perspectives of API Strategy. IEEE software.

# 6 API Lifecycle Analysis

## 6.1 Lifecycle Stage Characteristics

### 6.1.1 Intended Benefits:
- Identify the lifecycle stage of an API to aid in strategic planning and potential transitions
- Comparison across different APIs
- Look for inconsistency in API behavior, e.g., if an APIs is unstable but has support for many users, causing the organization to expend a lot of effort

### 6.1.2 When to use this method:
When it is not clear what lifecycle stage an API can be classified as, or whether the characteristics should be adjusted to better match strategic plans.

### 6.1.3 When not to use this method:
When the lifecycle stage of an API is already known and when the API behaves already as expected for an API of that particular stage.

### 6.1.4 Description
Each stage in the lifecycle identified in Section 6 can be characterized by considering how they fall within a range of characteristics, namely stability, change, commitment, governance, compatibility, and support. We summarize the mapping of each lifecycle stage to these characteristics in Table 3. These characteristics are summarized in the following.

*Table 3 Characteristics of each Lifecycle Stage*

|  | Plan | Operation | Deprecation | Retire |
| --- | --- | --- | --- | --- |
| **Stability** | Unstable | Mainly Stable | Stable | Stable |
| **Change** | Uncoordinated, Experimental | Coordinated, Change with cost | Minimal, Error correction | None |
| **Commitment** | No commitment to user | Committed to user | Decreasing commitment | No commitment to user |
| **Governance** | Setting up | Governed | Governed | N/A |
| **Compatibility** | No forward or backward compatibility | Forward and backward compatibility | No (API) forward compatibility | No forward or backward compatibility |
| **Support** | Intense feedback from few users | User support for many users | Minimizing support | No support |

**Stability** refers to the technical stability of the API. In planning stages of the development there is less stability and even fundamental design considerations can be revisited. When the API reaches pre-determined stability requirements and passes required testing, it is rolled out and acquires (hopefully) a growing user-base.

**Change** refers to the amount and type of change in each stage.  In the planning stage, change is uncoordinated and experimental.  In operation, change is coordinated and comes with a cost.   In deprecation, there is minimal change for error correct, and no changes typically occur in retirement.

**Commitment** means commitment to existing users of API. In an early stage there is no commitment to specific users since the API is not published yet. When the API is published, there is an immediate need to acquire users and develop the API based on user feedback. Later in the API life-cycle there is decreasing commitment to the users until ultimately APIs are retired and users need to find new substitute APIs.

**Governance** begins during planning, when the governance process is designed.  An API is governed in the operational and depreciation stage, but is no longer controlled during retirement.  More information on governance can be found in Section 8.1.

**Compatibility** refers to compatibility between the different versions of the API.   Generally, the usage SW is created with a particular API version in mind.   Wikipedia defines forward and backward compatibility as follows:

> "Backward (downward) compatibility is a property of a system, product, or technology that allows for interoperability with an older legacy system, or with input designed for such a system"

> "Forward compatibility or upward compatibility is a design characteristic that allows a system to accept input intended for a later version of itself."

If a usage SW client is intended to be compatible with a version of the API, but is also compatible with the next version of the API, from the perspective of the API, this is backwards compatibility.  From the perspective of the usage SW, this is forwards combability.   See Figure 8 for a visual description.  If a usage SW meant to be compatible with V2 of the API, but also works with V1 of the API, but is also compatible with V1 of the API, from the perspective of the usage SW, this is backward compatibility, but from the API perspective, this is forward compatibility.  See Figure 9 for a visual description.  A similar description can be made between the compatibility of the API and its implementation.

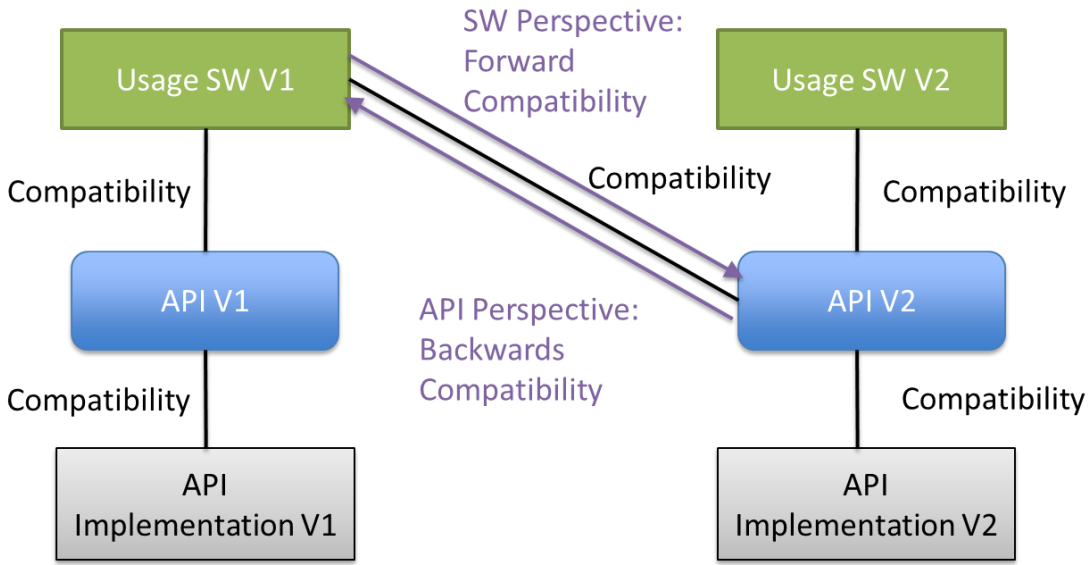

*Figure 8: API Backward Compatibility Example*

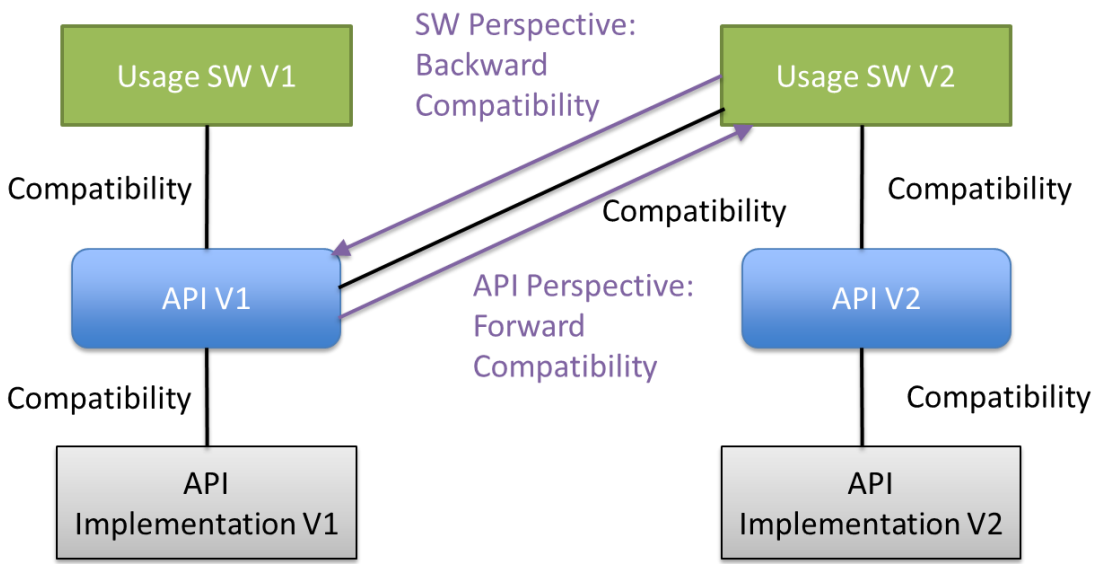

*Figure 9: API Forward Compatibility Example*

In the planning phase, the compatibility requirements do not yet have to be fulfilled, but they become extremely important when the API is launched: in this case, from the perspective of the API, it must be compatibility with clients in both a forward and backward direction. When the API is depreciated, the API forward compatibility becomes less important and ultimately both become unnecessary when the API is retired.

**Support** means the requirement to support the API. In the planning stage there is no existing user-base, so support is limited to preparing the users and managing expectations concerning the future launch. The need for user support is largest immediately after the launch and decreases when moving towards the end of the life-cycle and retirement.

### 6.1.5 How to apply

Consider the characteristics of an API of focus by going through the rows in Table 3, e.g., for my API, what is the level of Stability, Commitment, etc.    This can also be done for groups of APIs, classifying them in a Table, using the characteristics as a comparison point.

Forward looking questions:
If the characteristics of an API are different from what is suggested in the table, this deserves reflection.  For example, if there is an API that in the planning stage, but very stable, or an API in the operation stages, which undergoes many changes, this could be a sign of problematic design or planning.

If there are differences, are these differences deliberate?  Strategic?   If not, how can they be adjusted?

### 6.1.6 Examples

We've asked participant companies to describe how their APIs differ or remain conformant to the characteristics described in the table.  The results for three example APIs from three companies are shown below.  The description is for the following APIs; however, the characteristics are representative of several APIs in the companies.

      Device API:  API for an application to access the device specific settings.
      Technology API: General asynchronous API between application modules
      Network API:  Network APIs for configuring and operating video devices

| Characteristic | Device Settings API | Technology API | Network API |
|---|---|---|---|
| Stability | In the planning stage the API is mostly unstable but before going to operation is set to stable. During the rest of the stages is remains stable. | | |
| Change | | | Operation: High cost to change, often results in new APIs instead. Deprecation: No error correction done, unless security related. |
| Commitment | | All potential users participate during Plan since this is an internal API. | |
| Governance | The governance of the API starts before going to operation, meaning that the last part of the planning phase is the API Governance Process. | | Plan: Governance is predefined, the new API only has to be assigned to an existing category. |
| Compatibility | The API is only providing backward compatibility meaning an old application can run with the new settings service. If a new application is to be used the device needs to be updated (including the settings service) | | |
| Support | | During Plan and Operation we need to support all internal users. | |

*Table 4:    Lifecycle Characteristics for Example APIs*

### 6.1.7 Further Reading

Van Ommering, Rob. "Techniques for independent deployment to build product populations." *Proceedings Working IEEE/IFIP Conference on Software Architecture*. IEEE, 2001.

## 6.2 Strategic API Lifecycle

### 6.2.1 Intended Benefits:
- Allows API analysts and stakeholders to determine at which part of the lifecycle an API of interest lays. This helps to make strategic decisions about lifecycle decisions (e.g., is it time to go operational? Depreciate an API?)
- This also helps to have a common vocabulary in which to discuss the status of an API.
- The lifecycle framework also provides a way to compare and contrast the relative development of multiple APIs.
- Having an agreed upon lifecycle model upfront can act as a contract between API providers and consumers, e.g., providing rules and guarantees for deprecation and retirement.

### 6.2.2 When to use this method:
- When one is not sure about the maturity of a particular API.
- When one is planning a new API or change API, trying to come up with a methodology for API
- When one wants to compare API maturity.

### 6.2.3 When not to use this method:
- When the maturity and status of an API is already clear.

### 6.2.4 Description:
We have worked to understand the typical lifecycle stages through which an API progresses, including planning, operation, deprecation and retirement. We consider the lifecycle model as strategic as we consider business and organizational aspects when capturing API phases. We will outline and explain the model in more detail, the list a number of analysis questions one can apply to an API using ideas in the model.

The middle box focuses on the API itself, encompassing four major stages: Planning, Operation, Deprecation, and Retirement. The top green box focuses on the perspective of the Internal or External Customer and their software. The bottom grey box focuses on the perspective of the Internal or External Implementer of the API. Often the API designer and implementer are the same party, but not always. In each box there is a cycle of Build (& Update), Deploy, Measure, Learn (& Validate), a common way of capturing the lifecycle of software based on the Lean Startup Model. We include this cycle to emphasize the potential for frequent API updates and iteration, although the frequency of these updates will depend on the lifecycle state (Section 6.1). The Customer and Implementer perform various types of analytics in order to evaluate the usage or implementation of the API, and this feeds back into the measure and learn cycle. The side boxes represent the constant presence of change influencing the API, including the presence of new technology.

We describe each stage in more detail.

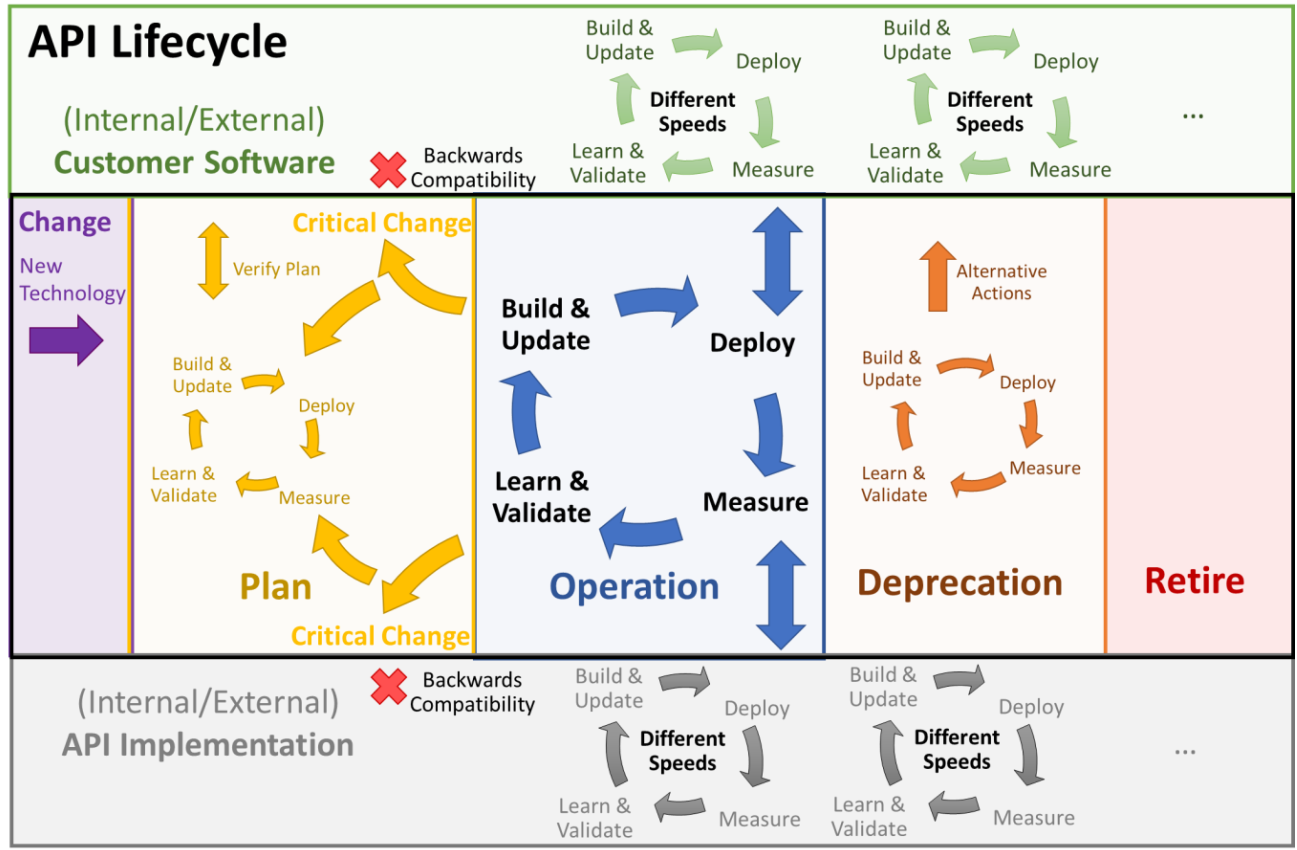

*Figure 10: Strategic API Lifecycle*

#### 6.2.4.1  Planning
In the planning stage, some form of market input helps to drive the need for an API.  The company envisions one or more business scenarios in which the API provides some internal or external value.  These plans and early designs could be shared with key customers to verify their utility, receiving key feedback.  When critical changes are made to an API sufficient to create a new version, the API can return back to the planning stage.

#### 6.2.4.2  Operation
Once the plan has been verified and the API has reached a sufficient level of maturity, the API is deployed and moves to the operation stage.  These transitions between stages are something we explore more in Section 6.3.

The build & update part of the cycle includes tuning, making smaller adjustments to the API.   The operation cycle has an emphasis on versioning, although versioning is also possible in the planning and depreciation stages, it is more critical to consider once the API has been released.  Versions could cover, for example, minor changes, more major changes, or critical changes.  In the last case, the API can be considered a new API, and the cycle returns to the planning stage, breaking backwards compatibility with the Customer Software and Implementation.  It is up to the organization to determine what is an is not a critical change, including who is allowed to make such changes.  Ideas for API governance are described in Section 8.1.  More minor or major changes are often in the name of performance improvements or for refactoring reasons, addressing technical debt.

Deployment can be in parallel, in that multiple versions of the same API could be simultaneously deployed or released, used by different parties.  In the measure part of the cycle, test cases are applied against the implementation, with results, as well as general feedback feeding into the learn and validate cycle.   This learning can provoke more changes, which may be classified by the size of their impact, e.g., minor, major or critical.

### 6.2.4.3    Deprecation

When the decision is made to retire or replace an API, it may go through a depreciation stage, where the API users can be prepared for the API retirement.   It is also possible to depreciate only part of an API (e.g., certain function calls or parameters) or to only retire certain versions of an API.  These decisions are likely driven by business change, and can also be influenced by new technology.  When an API is ready to be deprecated, it may still go through build, deploy, measure and learn cycles, albeit at a slower pace, and most likely only to fix important issues, not to extend the functionality.   In this stage, alternative actions are shared with the customers, instructing them on alternative APIs, if available, and preparing them for the eventual full retirement of the API.

### 6.2.4.4    Retirement

When an API is retired, it may be possible or desirable to retire only part of an API, or particular versions of an API.  In which case the remaining/parts versions are in previous stages of the lifecycle.  The process of retiring part of an API may be influenced by business change.

### 6.2.5   How to Apply

The general idea is to place the API of interest in one of these stages, and use the API lifecycle concepts for understanding and planning.   The figure represents the ideal lifecycle, but often there are departures from the ideal in reality.  The idea is to understand and mitigate these departures whenever possible.

Steps:
Identify an API of interest.   For this API, answer the following questions
- In what stage does the API fit?  Why?

The answer places the API in question in one of the four stages.   If there is difficulty in mapping an API to one particular stage, refer to Section 6.1 to explore the lifecycle characteristics in more detail.

If the API is in the planning stages, answer the following:
- What is the motivation behind the API creation?  In what business scenarios will it be used?  Does new technology facilitate the API creation?
- Are there critical stakeholders with whom we should verify the API plan?
- How do we collect feedback on the API?  From whom?
- Who are the API users and implementers?  Are the API implementers the same group/people as the API designers?  Different groups?
- Can we gather analytic/test data from the customer and implementer?  How?

Questions which are difficult to answer may indicate that there is some strategic area of the API that needs to be worked out in the organization.

Particularly for those APIs in the operation stage, answer the following questions:
- What change necessitates API Change?  Is this business change or a change in technology?

- What is the versioning scheme? What degree of change counts towards what type of version change (e.g., major, minor)? Are guidelines for these versions clear and consistent?
- What types/severity of changes can be made before the API is considered a new API?
- Is a critical change important enough to justify a break of backwards compatibility?
- Do changes which only enhance the performance, and don't change outward functionality count in the versioning scheme?

Answering these types of questions will help to compare operational APIs internally within a company and possibly externally.

If the API is in or near deprecation:
- Is a depreciation stage for an API necessary? Or should the API be directly retired?
- What are alternative actions when an API is deprecated?

If the API is in or near retirement:
- Should the entire API be retired, or should only part of the API be retired through refactoring?
- Does API retirement mean removal of API access, or only a complete removal of support and updates?

If time allows, it may be useful to look ahead to plan futures stages of the API. For example, if the API is in the planning stage, consider the questions for the operation, depreciation and retirement stage, i.e., how will versioning work, will there be a depreciation stage, at what point will the API be retired?

Forward looking questions:
Reflecting on the previous answers, are these elements which are problematic or need to be adjusted? Does the transition of APIs through the lifecycle match strategic needs?

### 6.2.6  Examples

Although we mean for our lifecycle model cover a variety of cases, in practice, implantation of API lifecycles differs between companies. Here we include examples provided by the companies of how additional APIs within their organization implement the lifecycle figure.

| Lifecycle Stage | | | | Layers | |
| --- | --- | --- | --- | --- | --- |
| Planning | Operation | Depreciation | Retirement | Customer | Implementation |
| Based on the requirements for the applications we identify what settings are needed in the device. During initial design of the device, the API version 1 is designed. The API design is then initially approved by the architect. The API is still in draft status.<br><br>Next step is that the API provider is implemented, and in the same iteration some applications are implemented that uses the API.<br><br>In the iteration the API is going through the API Governance process and is being approved. When it is approved it is set to status released (version 1). | In coming iterations the applications and the device gets new requirements. The API is then extended with new functionality (to version 2 ...) in a similar way that the version 1 was designed. (Draft, API Governance, released). The version 2 of the API is backward compatible with version 1 so not all applications needs to be updated.<br><br>The usage of the API is measured at the same time as the device as a whole is measured. (For usage, performance, etc.) | To enhance the system security a new API is created. All applications should migrate to use this new secure API. The API is set to status deprecated with a description how to migrate to the new API and during the following iterations, the applications are step by step migrated.<br><br>When building the system the build log is giving a warning that a deprecated API is used to inform developers not to use it any more. (And making sure no new API usage is created) | When there is no more measured usage of the API it is removed from the system and the settings service implementation removes the code in the next iteration. | The API users are the applications. (An application in this context is the UI of the device). These applications is sometimes made by us and sometimes made by other companies. | The implementation of the API is the settings service, running in the device. This is also done by us. |

*Table 5: API: API for an application to access the device specific settings (Device API)*

| Lifecycle Stage | | | | Layers | |
| --- | --- | --- | --- | --- | --- |
| Planning | Operation | Depreciation | Retirement | Customer | Implementation |
| This API was originally designed during around 2011 when the system was divided into function modules. It is used together with other API's when allocating resources. One function module is provider and several other function modules are consumer of this API. | Updated separately for each new feature that affects the API. High level proposal at feature pre-study and detailed design and test during feature execution. New protocol revision created for each new change. Different consumer function modules may use different protocol revisions in run-time. | Protocol revisions are deprecated when there exist later protocol revisions but not all function modules have upgraded yet. A discussion has started to deprecate this interface to start over with a fresh design. We seldom deprecate interfaces. | Protocol revisions are removed when no function module uses them anymore. We seldom remove interfaces. | Used by several independent function modules that different organizations within the company is responsible for. | Implementation follows design rules. The interface is defined using a text-based syntax/semantic. Code can then be generated from the definition. The generated code includes the API. Other frameworks are called from the generated code to handle communication on different platforms. |

*Table 6: General asynchronous API between application modules (Technology API)*

| Lifecycle Stage | | | | Layers | |
|---|---|---|---|---|---|
| Planning | Operation | Depreciation | Retirement | Customer | Implementation |
| A new API is ordered by a product owner, to support a new feature in the product or a new use case. The design and implementation of the new API is done by the team developing the product or feature and is usually iterated over a couple of months. A governing body of API architects provides guidance and feedback on the design decisions. | Initiated by publishing the finalized documentation of the API. Updates are made by the product development teams with input from API architects. Most APIs stay in this stage for at least 10 years. Updates are triggered by client feedback and/or change in hardware capabilities. | Sparsely used, due to poor overview of the utilization of most APIs, especially old legacy APIs where we have no well-defined use-cases. If we think an API is still in use, we continue support no matter what. When used, tend to stay in deprecation for many years. | We very seldom retire any APIs. Like deprecation, not much used, and for the same reasons. | Both internal and external users, but the big value come from external users. Users are typically integrators who integrate many different products from different vendors in their solution. | The API implementation is done by the same teams that design the APIs and maintain them. |

Table 7: Network APIs for configuring and operating Video devices (Network API)

### 6.2.7 Further Reading:
Ries, Eric. *The lean startup: How today's entrepreneurs use continuous innovation to create radically successful businesses.* Crown Books, 2011.   http://theleanstartup.com/book

## 6.3 Lifecycle Transition Points

### 6.3.1 Purpose
To identify, understand and evaluation transition points between API lifecycle stages.  To strengthen the strategic governance related to APIs by establishing formal decisions points related to transition points.

### 6.3.2 Intended Benefits
- Promote awareness of lifecycle stage transition points
- Improving transition points between lifecycle stages
- Consider governance of transition points

### 6.3.3 When to use this method
When transitions in the lifecycle are not clear or are not working as well as hoped.

### 6.3.4 When not to use this method
When the transition plans and procedures for an API are already established and unproblematic.

### 6.3.5 Description

We can consider the transition points between API lifecycle stages.  We have identified potential practices from our case companies as listed below.

Transitions to planning
- New product, project, platform
- Change in architecture
- Technology change or needs
- Business needs
- New security case, support same use case with different implementation
- Business case 1: very specific need and use case
- Business case 2: new case to promote and involve partners, prototyping and exploration
- Business case 1: short planning phase moving into operation with new release, the planning phase can be short compared to the operation phase
- Business case 2: releasing alpha or beta versions, eventually partners expect stability

Transitions from Planning to Operation
- When MVP is available, move fast to market
- Pre-release to selected customers
- Alpha and beta releases become implicitly operational over time
- Meets certain quality criteria – balance between quality and customer pressure
- Have a contract and freeze it, if people invest, they promise not to break the contract
- Public APIs can be standardized before going to market/becoming operational, this can be slow, internal APIs are faster to operation

Transition from operation to deprecation
- Cost of maintaining > value
- Marketing
- Technical debt
- Branch out, phase out platform
- Try to kill the market for the product, force users to buy new product
- Want to force new version, the new version is better, easier, more features
- Have a new version available, lower version depreciated, force users to switch
- In other cases, no new versions available. When service is not used, not economical, no wins
- Data-driven development, analytics provides data which indicates cost, usage and potential depreciation
- Some external cases are very hard to depreciate, internal cases are easier
- Discontinue support on market (but still exists)
- Provide 18 month warning/provide 2 release warnings
- Give time to move to new API

Transition from deprecation to retirement
- Park development track and associated APIs
- Security issue forces immediate retirement of API
- When depreciating, decide how long depreciation stage is (this decides retirement)
- Move to deprecation and retirement when a major version with breaking changes is released
- Aim for 6 month warning between deprecation and retirement (management has to be on board)

- Many companies have problems deprecating and retiring APIs in practice.   This is an issue related to technical debt

### 6.3.6   How to apply

Inspired by the list above, for each API analyzed, identify the transition points between stages.

Forward looking questions:
- Are these transition points reasonable?  Desirable?
- Are there other reasons to transition?
- Is there governance or control or some guidance on making such transitions?  Should there be?
- How could or should the transition criteria change?

#### 6.3.6.1   Lifecycle Value analysis

One can consider the value lifecycle of a feature, as in the figure below.  Similarly, we can consider a value curve for APIs, trying to understand at what point the value increases, maintains, and decreases.

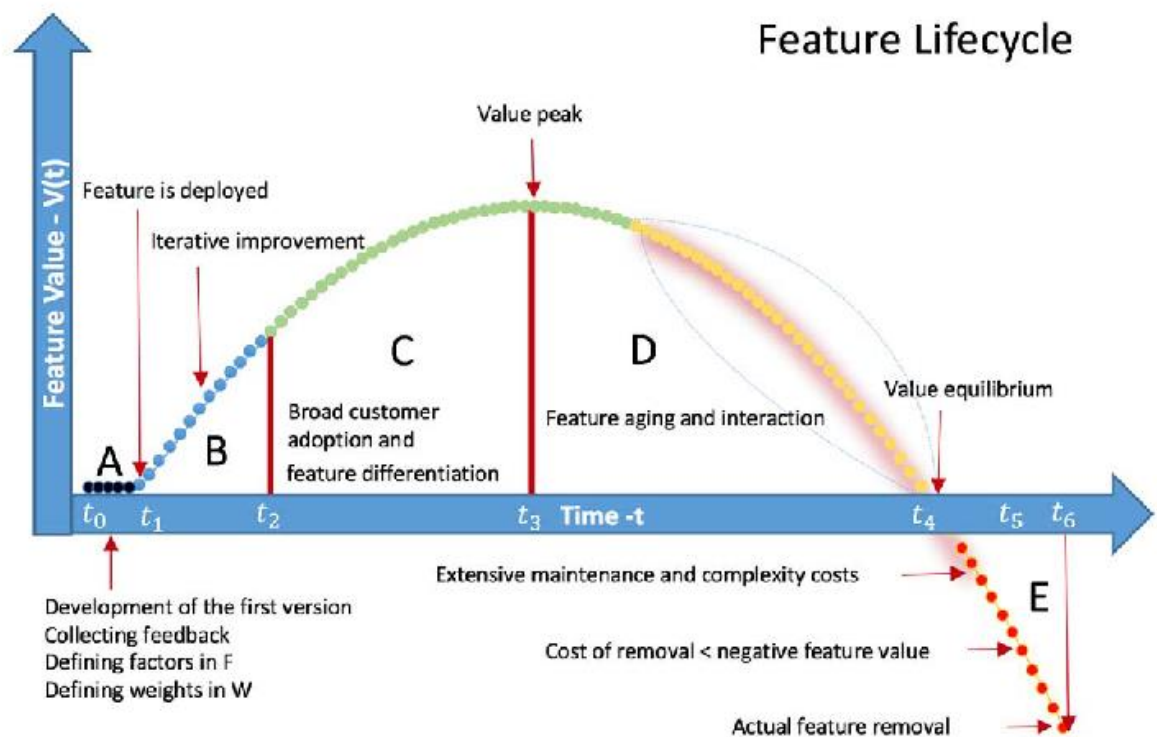

Figure 11:  Feature Lifecycle Model from [Fabijan, Olsson, Bosch, 2016]

We want to map this value curve to the API lifecycle, as shown, for example in the figure below, looking for mismatches.

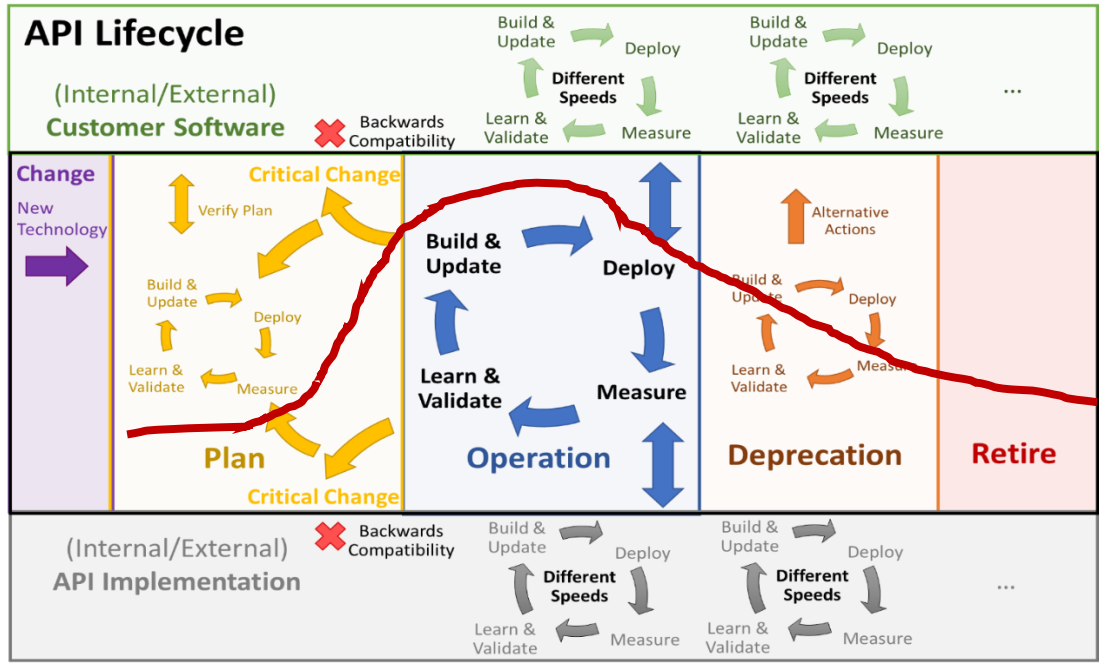

*Figure 12: Visualization of API Value over Lifecycle*

Possible mismatches to look for:
- The value of an API is very high in the planning stages
- The value of an API is not on the rise before operation
- The value of an API is never very high, yet operation is reached
- The value of an API decreases greatly while still in the operation stage
- The value of an API is still very high when moving to deprecation or retirement

If any of the above cases apply, consider if the transitions are not happening at the optimal times, or if the APIs are classified or moved to the wrong stage.

# 7   API Analysis with Modeling
## 7.1   Value Modeling

### 7.1.1   Intended Benefits
- Provides a high-level view of the API ecosystem, provoking shared understanding and discussion.
- Allows for a comparison between APIs, internal or external to a company.
- Includes both technical and non-technical components.
- Understand motivations for APIs at a high level.
- Understand the values provided by an API, and to whom the values are provided.
- Understand missing or problematic value flows.  Those that are desired but not yet completely fulfilled.

### 7.1.2   When to use this method
When one needs to understand the actors in an ecosystem and how they provide and receive value.  To understand the reciprocity of exchange, and why various actors participate in the API ecosystem.

### 7.1.3 When not to use this method

When the API ecosystem, actors, and exchange of value is clear. It is well understood what benefits every player receives in the API ecosystem.

### 7.1.4 Description

Given the competitive nature of business, technology development and strategy has evolved to focus on value [1, 2]. This focus aims to ensure that technology, processes, or interactions have a clear contribution to customer or business value. Furthermore, the emphasis on value as part of agile methods, as well as the focus on value in, for example, value-based software engineering [3], has provoked a recent academic focus on value analysis and modeling. Several value-oriented modeling approaches have been introduced; in this Section we use e3 value modeling as per Gordijn et al. [4]. We select this language due to its simple visual syntax, continued application, development in research (e.g., [5]), and availability of tool support.

We describe the basic concepts of e3 value modeling in the following.

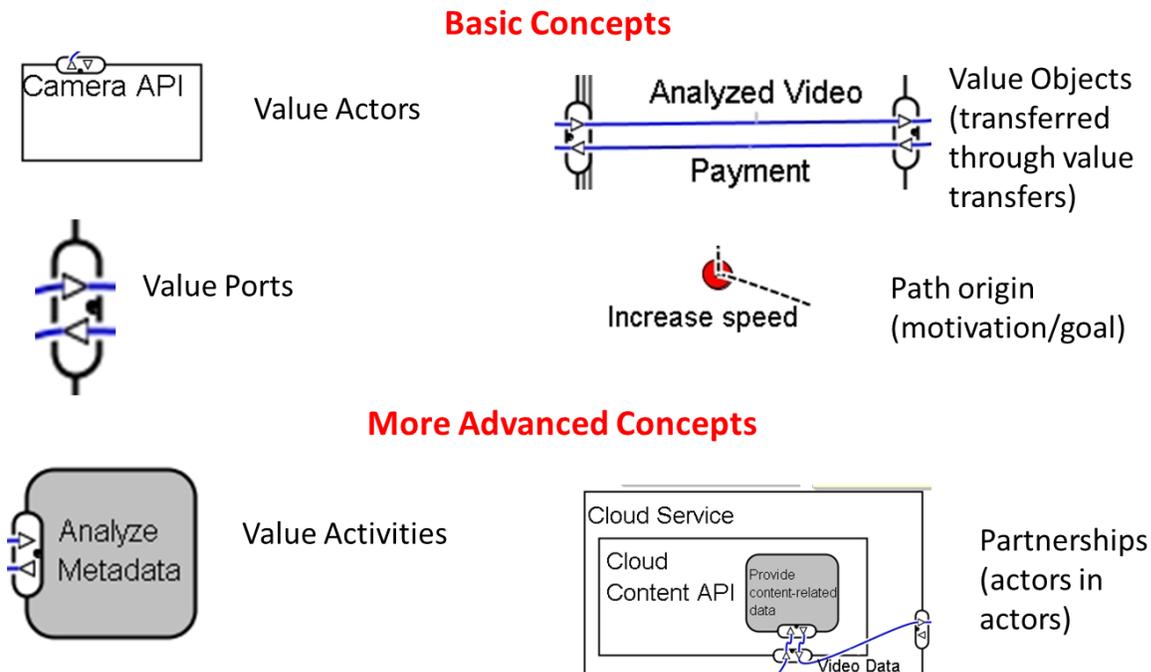

*Figure 13: e3 Value Model Concepts, Simplified from [4]*

Actors: these are the actors in the value ecosystem. They can be a company, a role (e.g., customers), a specific group or person. Although they were intended to be used for people-oriented entities, we have also used them to represent technical actors such as APIs, servers, and architecture elements.

Value flows and objects: actors exchange value objects with each other through value flows. One actor provides value to another, and usually this actor must get some value back, either directly or indirectly. A value can be something concrete, technical or abstract, e.g., data, functions, capabilities, payment, strategic advantages, market share, trust, or knowledge.

Value Ports:  value flows flow in and out of an actor through ports (triangles).  They can be grouped into interfaces (black oval), usually to show that the value flows are related, perhaps an in and out exchange.

Path origin:  value models can contain flow, including a start and stop stimuli.  The start stimuli (red circle) can be used to indicate the start of a value flow, or the motivation or stimuli.

Value activities:  inside of an actor, activities can produce the exchanged values.   Value can flow from activities inside of actors instead of the actors themselves.

Partnerships:  an actor can be inside of another actor, e.g., an employee or role inside of a company.

### 7.1.5   Examples
Through our work with our partner companies, we have developed several e3 value model examples.  We show several of them in the following for inspiration.

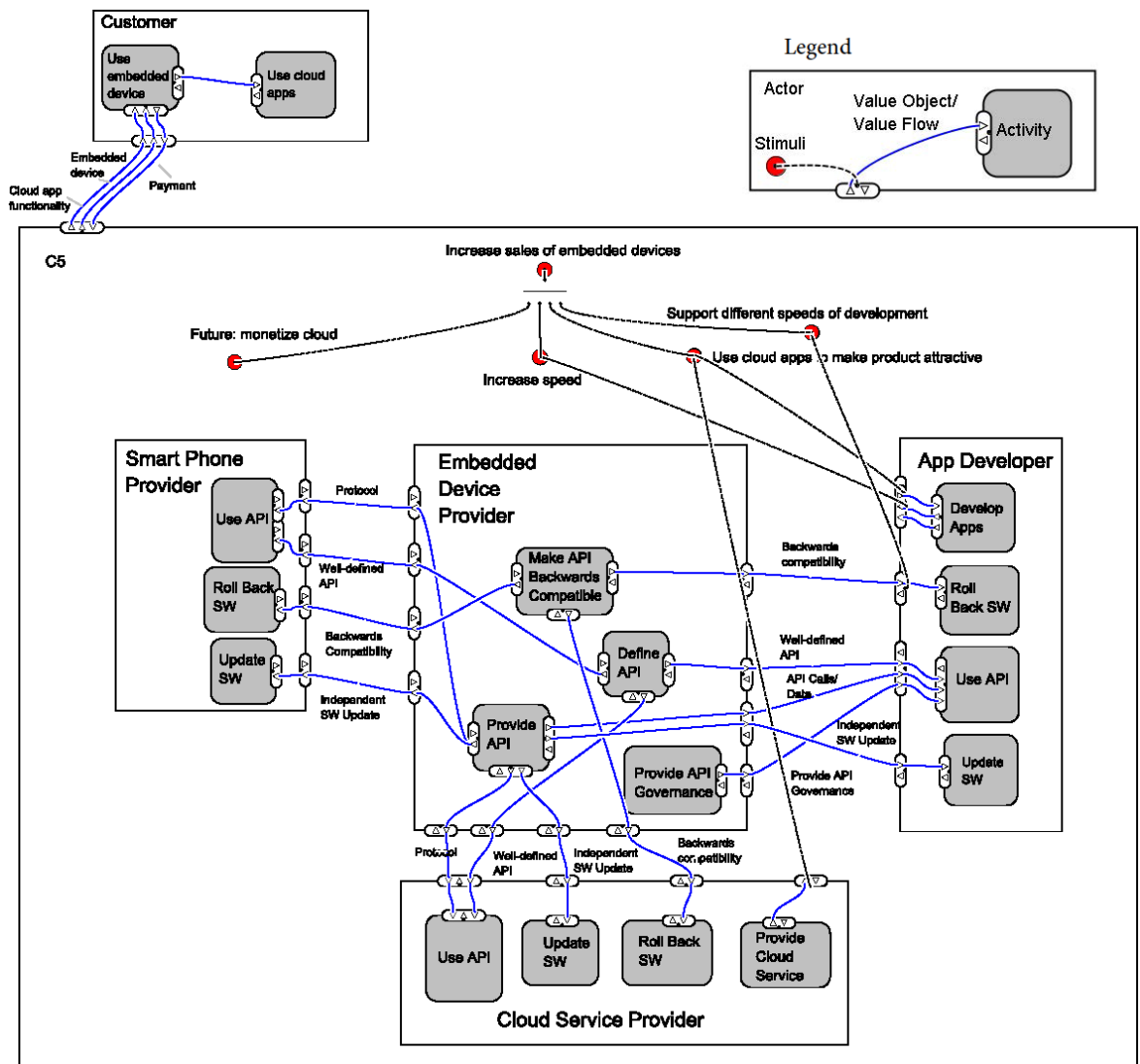

*Figure 14:  Device API e3 value model example*

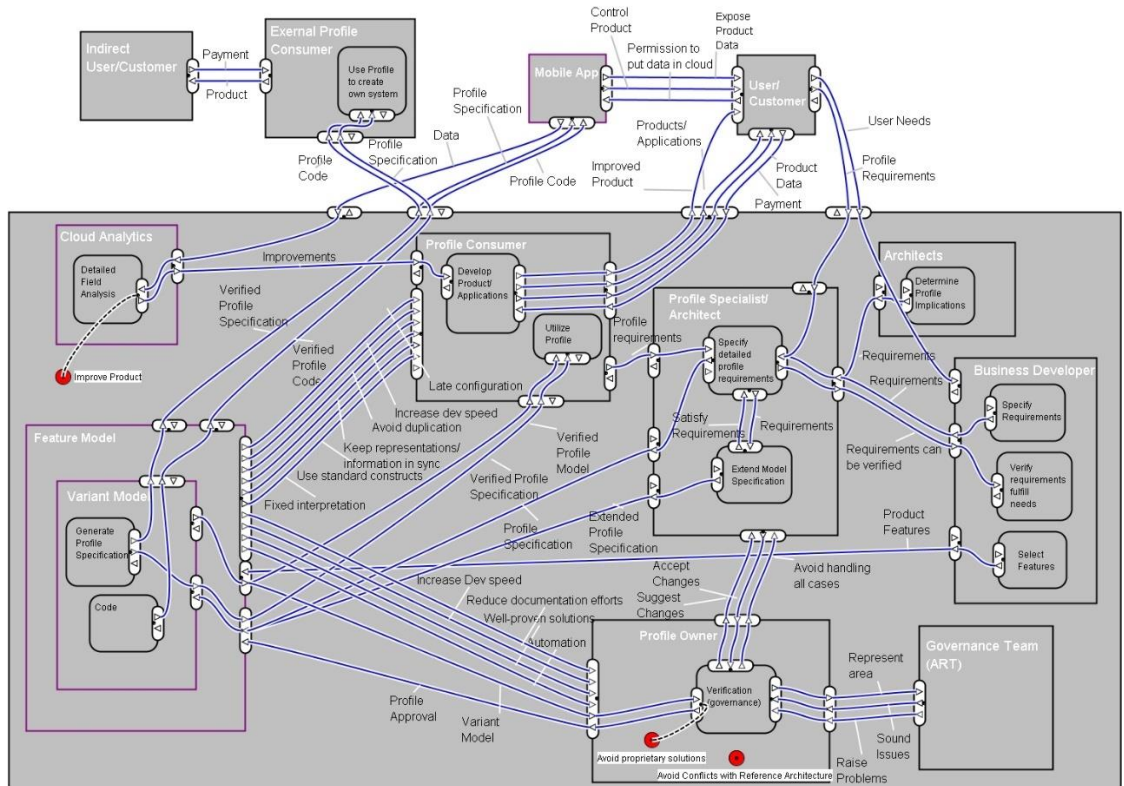

*Figure 15: Product API e3 value model example*

### 7.1.6 How to Apply

We provide a set of steps to guide one or more people through the value modeling process.

**Model Driver.** Our past experience have indicated that these sessions are more successful if a driver or moderator directs the session and the modeling. Ideally this person has some past experience in value modeling, or modeling in general, and can share their knowledge. However, in the case when this experience in lacking, we recommend that the model session driver thoroughly reads the full set of method guidelines, while the other participants can view a shorter summary. We provide a simpler set of steps for all participants, than a more detailed set of steps for the facilitator or driver of the session.

**Value Mapping Method – High-level (for Participants)**

The steps are a recommendation, the ordering can be adjusted as desired.

1. Scoping: pick problem and limits – what is the focus? What is out of scope? Model as-is or to-be?
2. Create Draft value model, in any order:
    a. Identify and draw relevant actors – who is involved?
    b. Capture needs via stimuli  - what are we trying to accomplish?
    c. Identify and capture value flows  - what value to do they exchange? Go beyond data flows
    d. Consider value activities – what do the actors do to provide value? With the value?
    e. Consider reciprocity  - does everyone get something in return?

    f. Identify cross-cutting API  - where is the API?
    g. Analysis/Completeness checks – what is missing?
      i. (Optional) API layers – how can the actors map to the layers in Section 5.1.

**Value Mapping Method – More Detail (for Model Drivers)**

We describe more detailed steps if the above steps are unclear, or for the person who will be driving the modeling.  The steps are a recommendation, the ordering can be adjusted as desired.

Iterate over the following steps in order to create a draft value model:

1. Scoping: pick problem and limits
    a. Pick an API ecosystem of interest
    b. Continually watch for scoping and "scope creep" in the modeling process.  Question whether all new actors are relevant.
    c. Decide whether to model as-is or to-be, current or desired state.

2. Identify and draw relevant actors
    a. Who uses the API?
    b. Who creates it?   Changes it?
    c. Who governs it?
    d. Are actors part of other actors?
    e. Are actors instances or market segments?
    f. Which actors are internal/external?
    g. Possible extension: distinguish between human/software actors with different colors/shapes.

3. Consider needs/stimuli.  Why does the ecosystem exist?
    a. Consumer needs
    b. Internal company needs

4. Identify and capture value flows
    a. What values does the API provide?  To whom?
    b. Hint:  ensure to capture not only technical value flows like functions, but also desired qualities and strategic elements, e.g., market share, branding, trust.   Data flows are OK to start, but try to go beyond.
    c. Connect value flows to needs (stimuli)
    d. Possible extensions:  capture desired but missing value flows (e.g,. in red) capture problematic value flows.

5. Consider and capture value activities
    a. What is the activity that creates the values for each actor?
    b. Consumes values?

6. Consider reciprocity
    a. For each existing flow, is there a backflow?
    b. Does each actor both get and provide value?
        i. If not, why?  Is this justified?  Aligned with needs?

7. Capture the API (if not already present)
    a. Is it value flow or interface?
    b. An activity?
    c. One or more actors?
    d. Could be cross-cutting across many value flows.

8. Pruning/refinement
    a. As individual or in workshop (in workshop if not previously)
    b. Is every actor relevant to the API ecosystem?
    c. Is every value flow relevant to the API ecosystem?
    d. Is the wording of actors/flows clear
    e. Is the overall motivation for the API clear?
    f. Is it clear for every actor?

9. Completeness
    a. Identify governance aspects in the model
    b. Are APIS Layers covered, can you map every actor to   Asset, API, Usage, Domain?
    c. Are BAPO elements covered:  business, architecture, process, and organization?

### 7.1.7 Tool Support

Whiteboard or pen and paper can be used.   Online tools exist, but are not simple to use in a collaborative setting (https://research.e3value.com/tools/).  They could be used offline later to capture the model.   Other common drawing tools like Draw.io or Omnigraffle can also be used.

### 7.1.8 Further Reading:
1. Miller, P.: Explaining technology's value. Forbes (May 2015)
2. Kiessel, A.: How strategic is it? - assessing strategic value. Oracle (February 2012)
3. Biffl, S., Aurum, A., Boehm, B., Erdogmus, H., Grunbacher, P.: Value-based software engineering. Springer Science & Business Media (2006)
4. Gordijn, J., Akkermans, H., Van Vliet, J.: Designing and evaluating e-business models. IEEE intelligent Systems 16(4) (2001) 11-17
5. Gordijn, J., Petit, M., Wieringa, R.: Understanding business strategies of networked value constellations using goal-and value modeling. In: Requirements Engineering, 14th IEEE Int. Conf., IEEE (2006) 129-138

## 7.2   Ecosystem Modeling with Goal Models

### 7.2.1   Benefits

The ***benefits*** of this method includes:
- An alternative visual view of an API ecosystem
- Articulating the business and individual goals of each actor in the ecosystem
- Capturing goal hierarchies
- Capturing and evaluating alternative strategies
- Linking interactions (dependencies) between actors to actor goals, capturing why

- A focus on API qualities or NFRs in an ecosystem

### 7.2.2 Use Cases

Possible use cases:
- Design of a product ecosystem for which the company has control
- Understand an existing or forming ecosystem to evaluate whether and in what ways to join the ecosystem
- Compare ecosystem alternatives
- Improve ecosystem interaction and design

### 7.2.3 When to use this method
- When an ecosystem is not clear, including how different actors interact and what is the benefit of these interactions
- When decisions about an ecosystem have to be made
- When ecosystem interactions should be improved

### 7.2.4 When not to use this method
- When an ecosystem is well understood
- When there is no control or choice in the ecosystems

### 7.2.5 Method: How to construct a goal model capturing an ecosystem

#### 7.2.5.1 Tooling
To draw a goal model, one can use paper or white board. However, this can often cause difficulties if the model evolves or changes as you draw. Other tool options:

http://creativeleaf.city.ac.uk/ (use in Chrome)
Cookies/data collection is option, at the moment the data is accessible only by Jennifer Horkoff

https://sites.google.com/site/creativitygm4re/quickinstructions

Alternative version: http://creativeleaf.portal.chalmers.se/ (Use in Chrome)

Downloadable older desktop tool (in Eclipse)
https://sourceforge.net/projects/openome/files/
Other modeling tools such as Visio or Draw.io might work, but then the goal-model specific template is missing.

#### 7.2.5.2 Actor and Dependencies

Step 1: consider and draw actors in the ecosystems. This includes agents and roles: companies, people, organizations or groups within a company, regulators, customers, users. This can also include technical agents like: a specific system, a cloud, an API, etc. Whatever has goals assigned can be an actor.

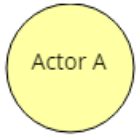

Identify dependencies between actors in an ecosystem. These can be resources or data, tasks, goals, or qualities, e.g.,

Actor A depends on actor B to provide resource R.

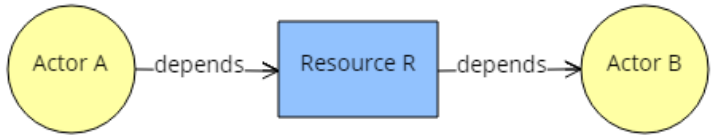

Actor A depends on actor B to do task T.

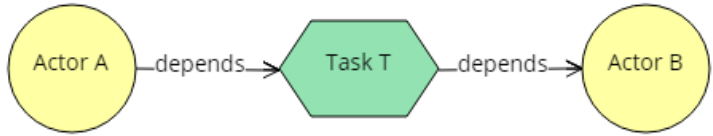

Actor A depends on actor B to satisfy or accomplish goal G.

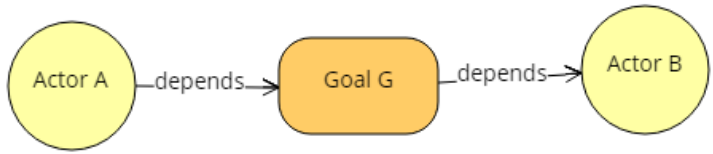

Actor A depends on actor B to sufficiently satisfy quality Q.

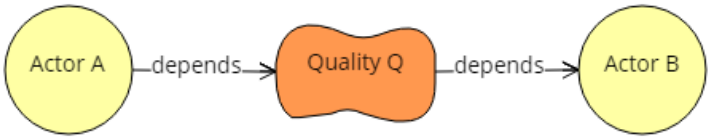

The model can be saved at this point, it is a simple view of an ecosystem.

### 7.2.5.3 Actor Rationale

In the previous section, we have identified who is involved in an ecosystem and what they depend on each other for. Now we would like to explore the rationale of each agent. Why do they participate in an ecosystem? What are their goals? Tasks? Qualities? Is there only one way for them to do things, or are there options?

Now we can "open up" each actor in the previous model to explore their rationale.

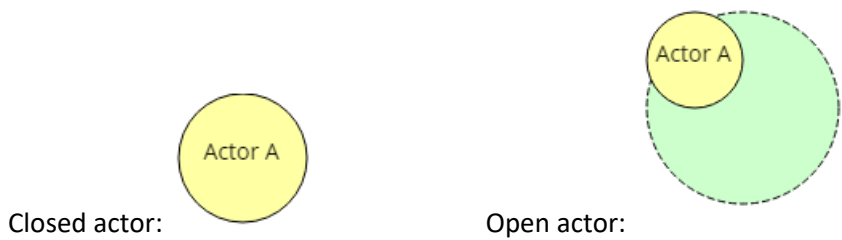

Closed actor:                    Open actor:

What are the goals of each actor?  Why do they participate in an ecosystem?  Goals can be functional or clearcut, or more qualitative.  They can be goals or quality goals.  Draw the major goals and desired quality goals of an actor in their boundary.  Try to do this for all the important actors.  Less important actors can be left closed.

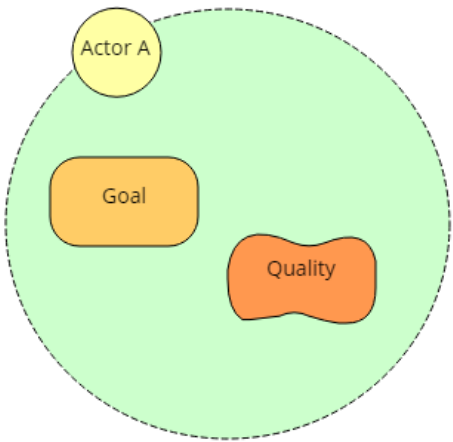

What are the tasks of an actor?  What do they have to do?  Add tasks.

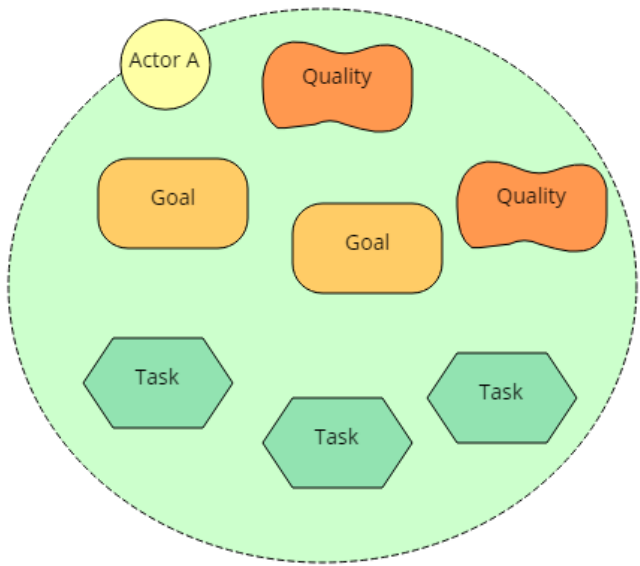

How do these elements inside an actor connect?  Add AND/OR links between goals and tasks (not qualities).  Note that this might lead you to find additional missing goals and tasks as you ask why? (moving up the model) and how? (moving down the model).

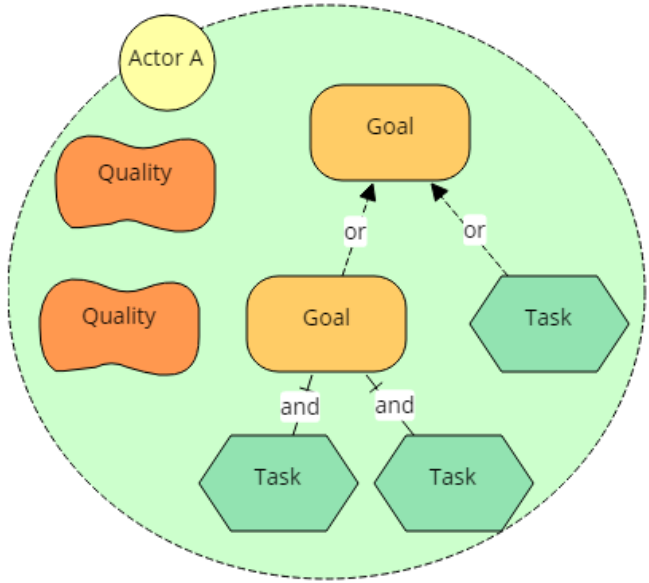

How do the goals and tasks effect the achievement of qualities?   Add positive and negative contributions to quality goals.   There are four types:  Makes, Helps, Hurts, Breaks, from strong positive to strong negative.

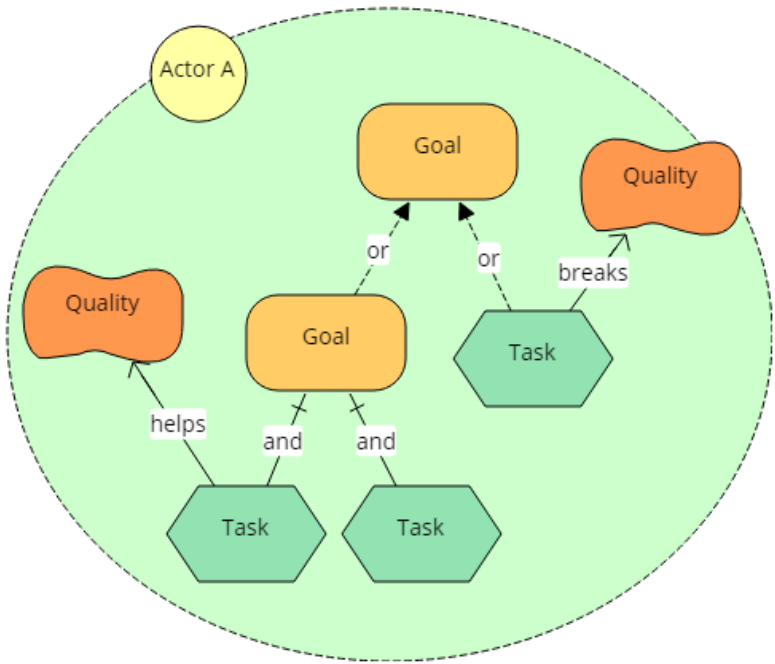

Finally, for the open, important actors, attach the dependency links found earlier to the goals, qualities and tasks inside of an actor. This answers "why?" and "how?" for each dependency link.   How do I provide this?   Why do I need it?   Usually there are dependencies going in both directions, e.g., Actor A depends on actor B, but actor B also depends on actor A for other goals/qualities/tasks/resources.

Completing this step shows how an actor fits or could fit into an ecosystem.

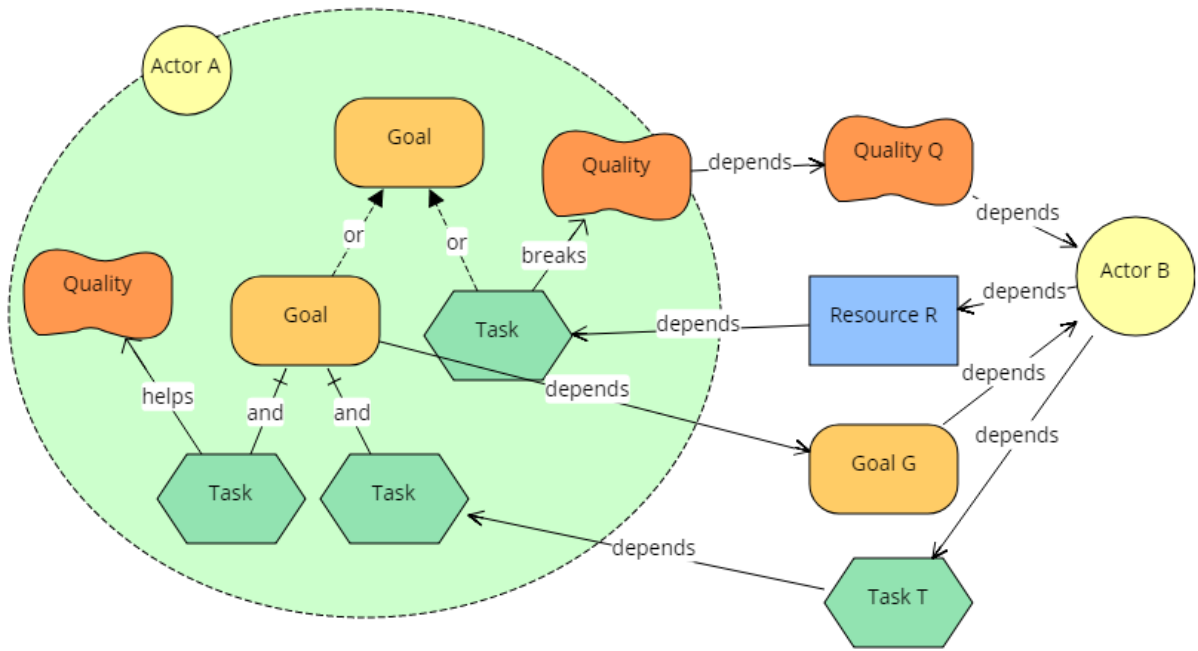

Note: Iteration. The above steps will likely be highly iterative. Connecting goals, tasks and qualities can cause you to think of new qualities. Connecting dependencies will help to find missing goals/tasks/qualities/resources, and even new actors.

### 7.2.6 Examples for API Ecosystems

*Reporting API* example comparing the rationale and actions of an actor in the near and farther future:

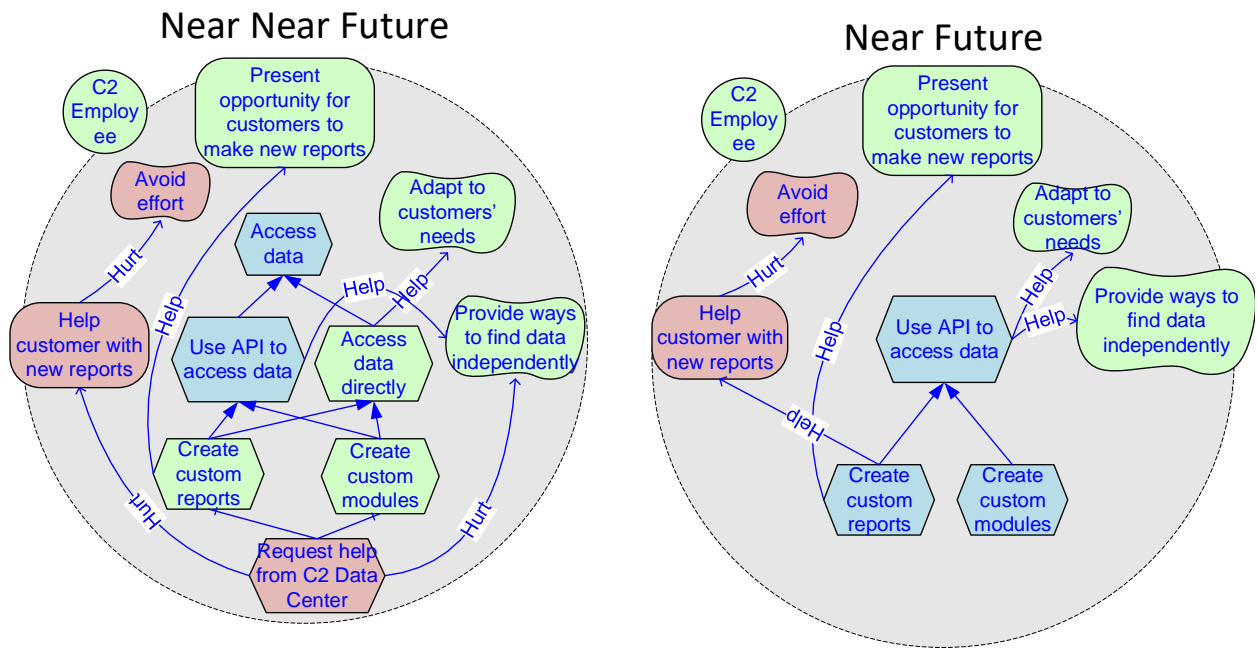

*Figure 16: Example comparing time snapshots using an API Ecosystem Goal Model, Detailed*

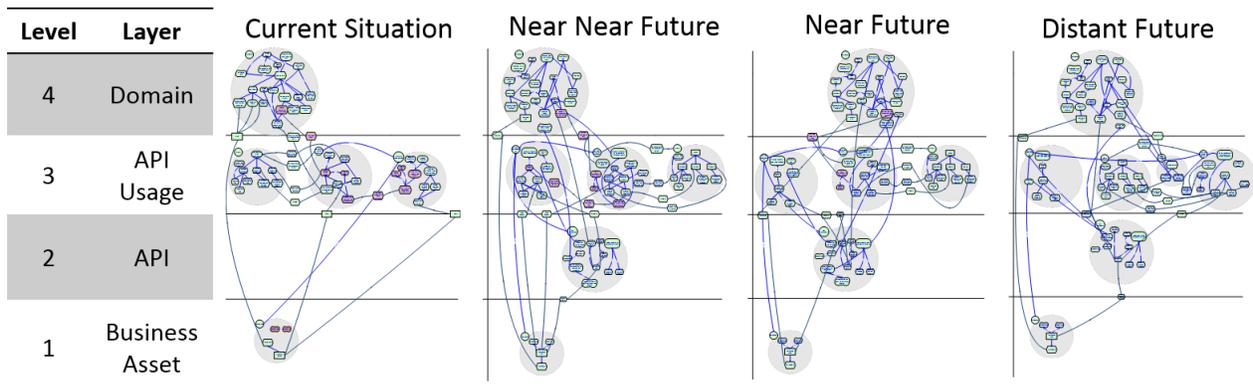

*Figure 17: Example comparing time snapshots using an API Ecosystem Goal Model, Overview*

### 7.2.7 Evaluating Health of the Ecosystem

We can use the models to assess whether the goals and quality goals of the actors in the ecosystem are satisfied, or to understand where the actors may be vulnerable to failing goals.

We use qualitative labels to capture goal, task, quality, and resource satisfaction.

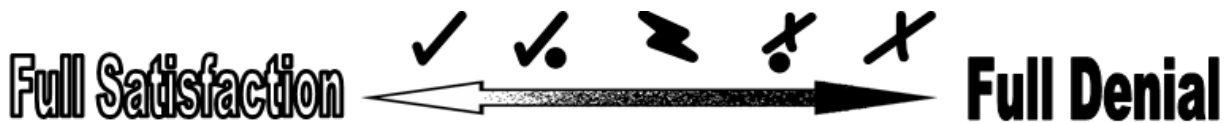

Propagation: one can propagate labels in the model using the meaning of the links, e.g.,

If a dependency is satisfied, the goal/quality/task that depends is satisfied.
If a goal/task is AND refined and its sub-elements are satisfied, it is satisfied.
If a goal/task is OR refined and one of its sub-elements is satisfied, it is satisfied.
If a goal/task helps a quality, it is partially satisfied.
If a goal/task hurts a quality, it is partially denied.
If a goal/task makes a quality, it is satisfied.
If a goal/task breaks a quality, it is denied.

One can also assess the satisfaction of the elements using knowledge of the ecosystems. Do you think the goal is satisfied right now? If yes, does the model show how? If not, why not?

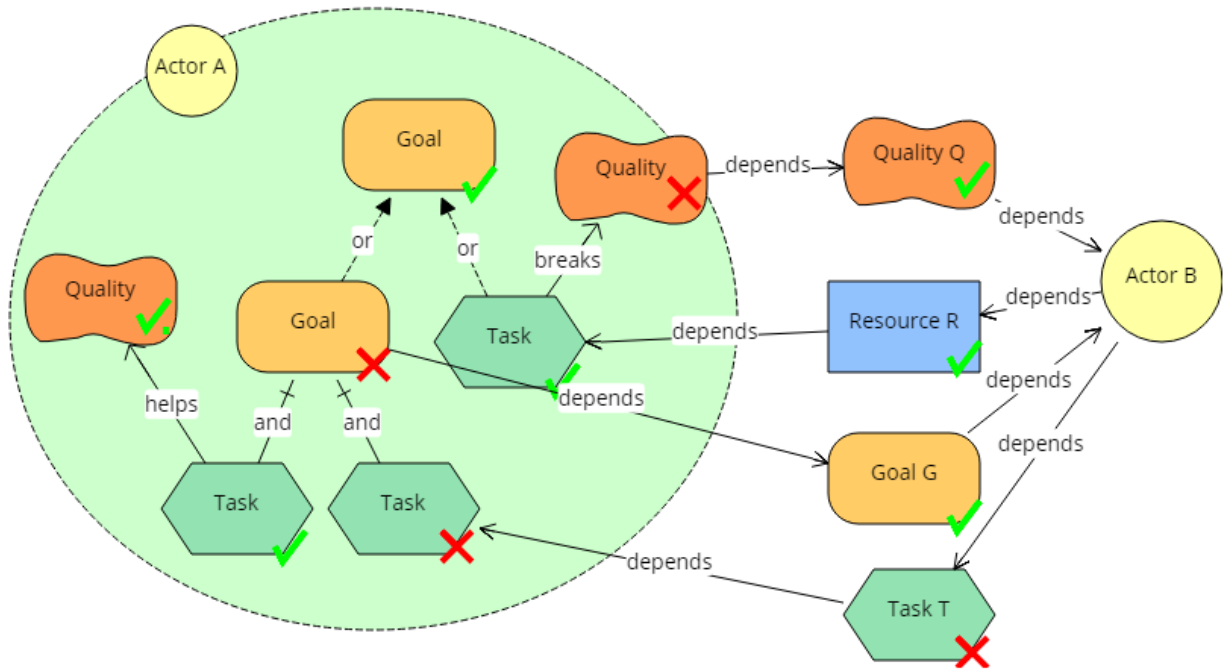

In this way, one can evaluate how well the ecosystem meets the goals of the actors. If an actor has many goals or qualities unmet, it is not successfully participating in an ecosystem. It may drop out, if possible.

In particular, focus on the goals and qualities of the company actors, is the ecosystem satisfying its goals? If not, why not? What can be changed?

### 7.2.7.1 Example
Figure 18 shows part of the Ecosystem Model for the Product API showing goal evaluation. Here we see problems in the satisfaction of qualities.

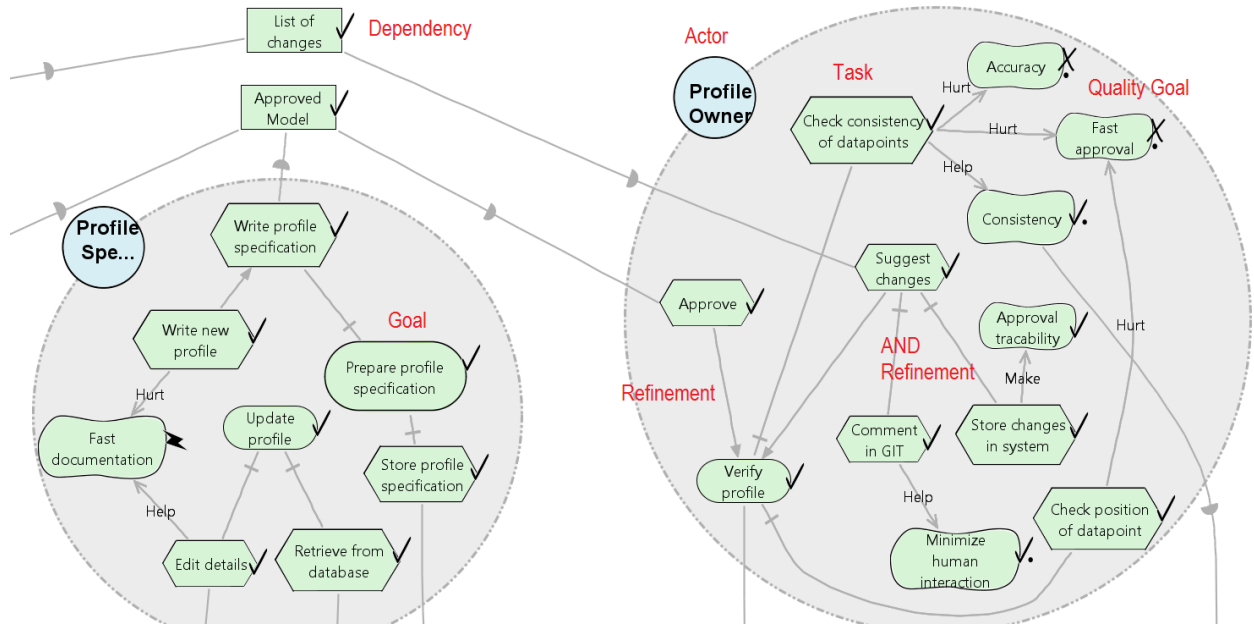

*Figure 18: An Example of Goal Model Evaluation for an Ecosystem Model*

### 7.2.8 Evaluating Ecosystem Alternatives

There are two ways to capture alternatives in goal models. The first is through the OR link. This is a lower-level alternative which captures more than one way of doing thing in the same model.

The other way is to draw alternative versions of a model. For example, if there are multiple ways for a company to join and interact with an ecosystem (connections shown via dependencies connected to goals, tasks and qualities), then there can be multiple versions of a model, each showing different ways to connect.

The different ways to connect will have different impacts on the goals of the actors in the ecosystem. For each viable way to connect to an ecosystem, consider how goals are met, particularly the goals of the company. Is there a viable way to connect? What is the best way according to desired goals and qualities?

One way to connect to an ecosystem:

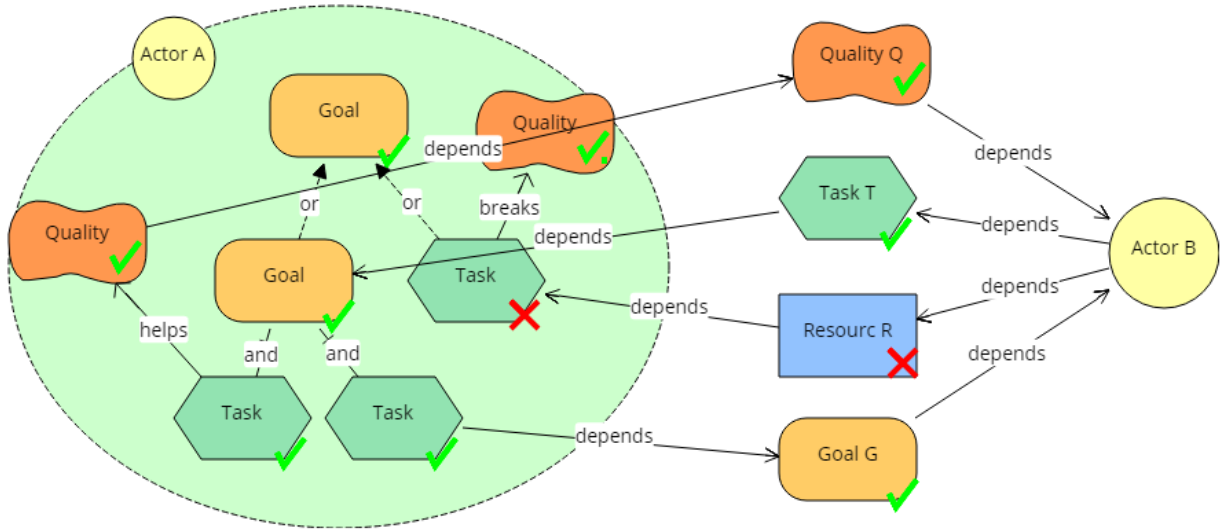

Another way to connect to an ecosystem:

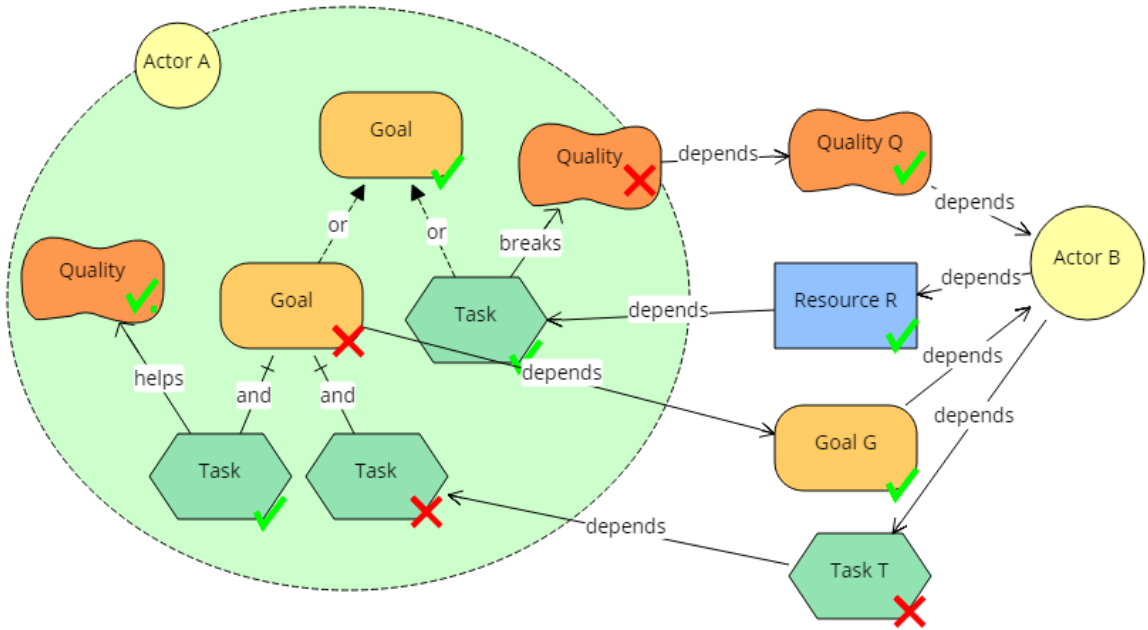

Generally, the top options is better for actor A, as more goals are satisfied. We do not know which one is better for actor B. In addition to connecting to an actor in multiple ways, we can also connect to different actors, components or systems.

## 7.3    Layering Goal Models

### 7.3.1    Intended Benefits
- Allows one to find missing critical actors in the ecosystem model
- Allows viewing of the API ecosystem using a layered view, classifying actors

- Manage the complexity of a large API ecosystem

### 7.3.2 When to use this method
When you have a goal ecosystem model and are interested in viewing the results using the API layers, classifying actors in terms of the API.

### 7.3.3 When not to use this method
When you have not yet created a goal model, or the goal model is clear.

### 7.3.4 Description

After a goal model has been created as in Section 7.2, one can split the model into layers using the API layers as in Section 5.1.   This helps to sort and classify  actors in the model, dealing with model complexity.

We recommend the following method:
- Consider each actor in the goal model and ask:
    - Is this actor capturing the API or it's components?
    - Is this actor capturing the assets protected by the API?
    - Is this actor capturing the SW using the API?
    - Is this actor capturing people or organizations in the domain of the API usage, i.e. this directly or indirectly using the usage SW?
- From these questions, place each actor into a layer.  Move the actors around on the canvas.  Add text annotation to indicate which layer is which.

See an example in Figure 19.   In this example, the first version of the goal model did not capture the API itself.  The model was revised to add new actors accordingly.

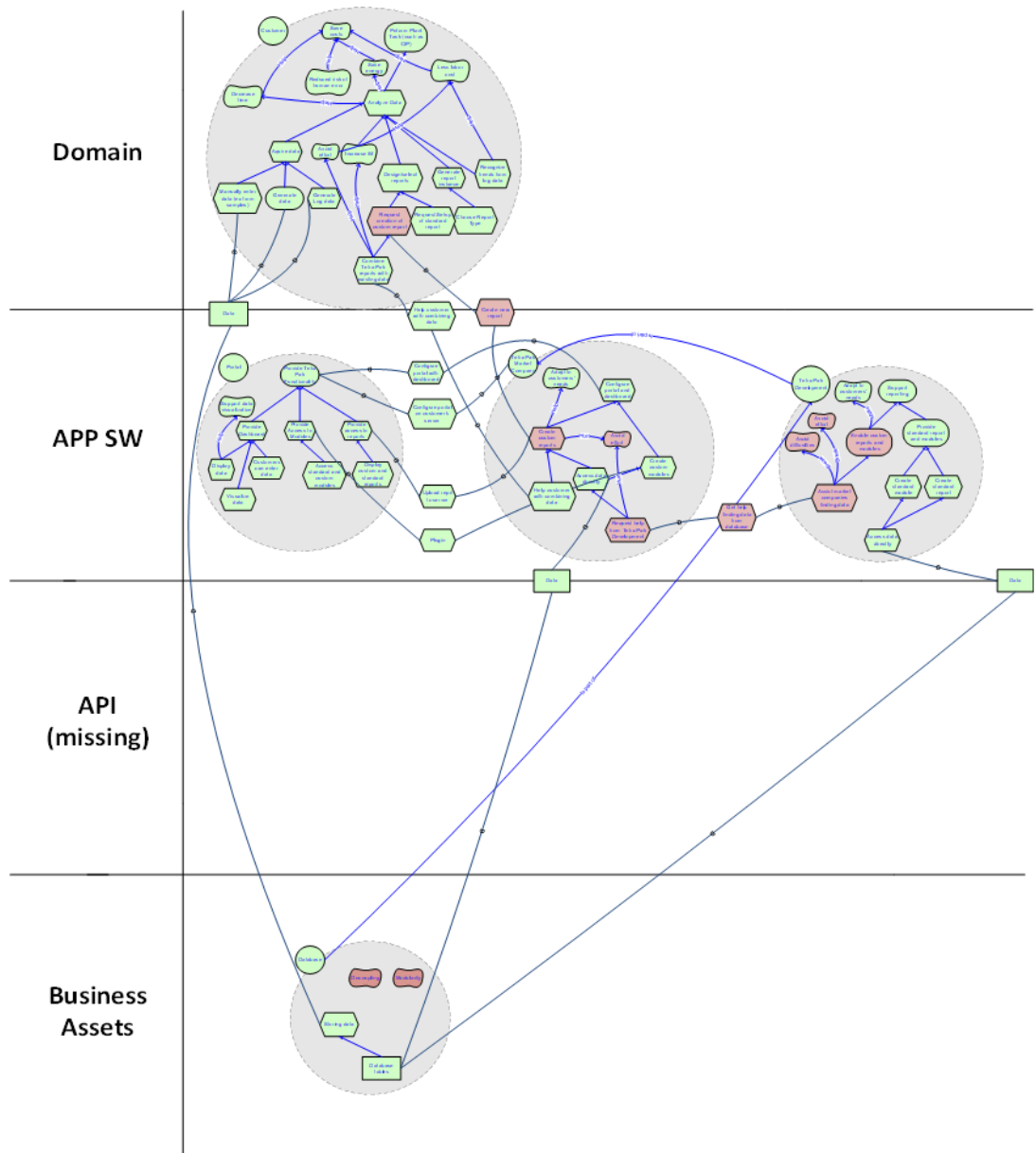

*Figure 19: Example layered goal model for the Reporting API*

Note that if the model contains more than one API, depending on the API of focus, the actors will be placed in different layers. A full, detailed example is shown in Figure 20, with a simplified version showing actors moving from layer to layer depending on the API of focus is show in Figure 21.

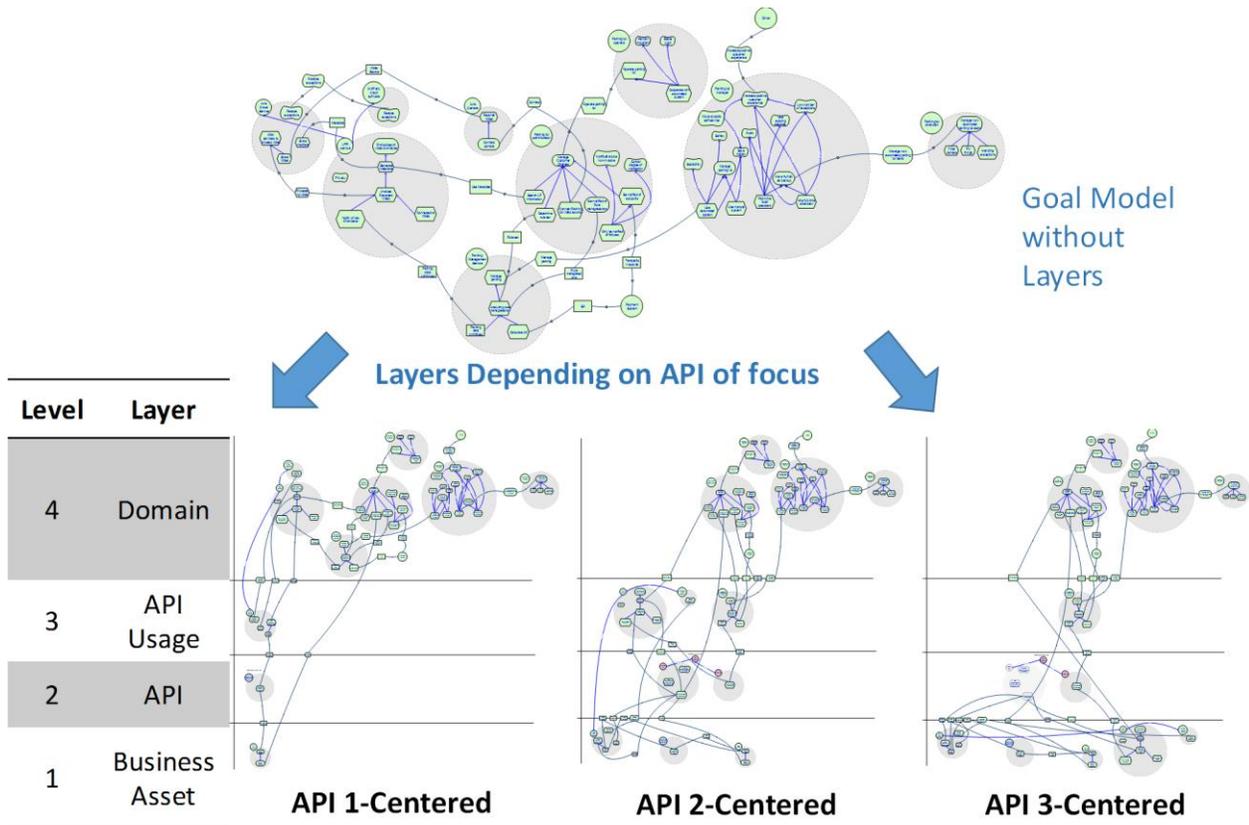

Figure 20: Ecosystem Model Mapped to API Layers depending on API of Focus, Full Example for the Reporting API

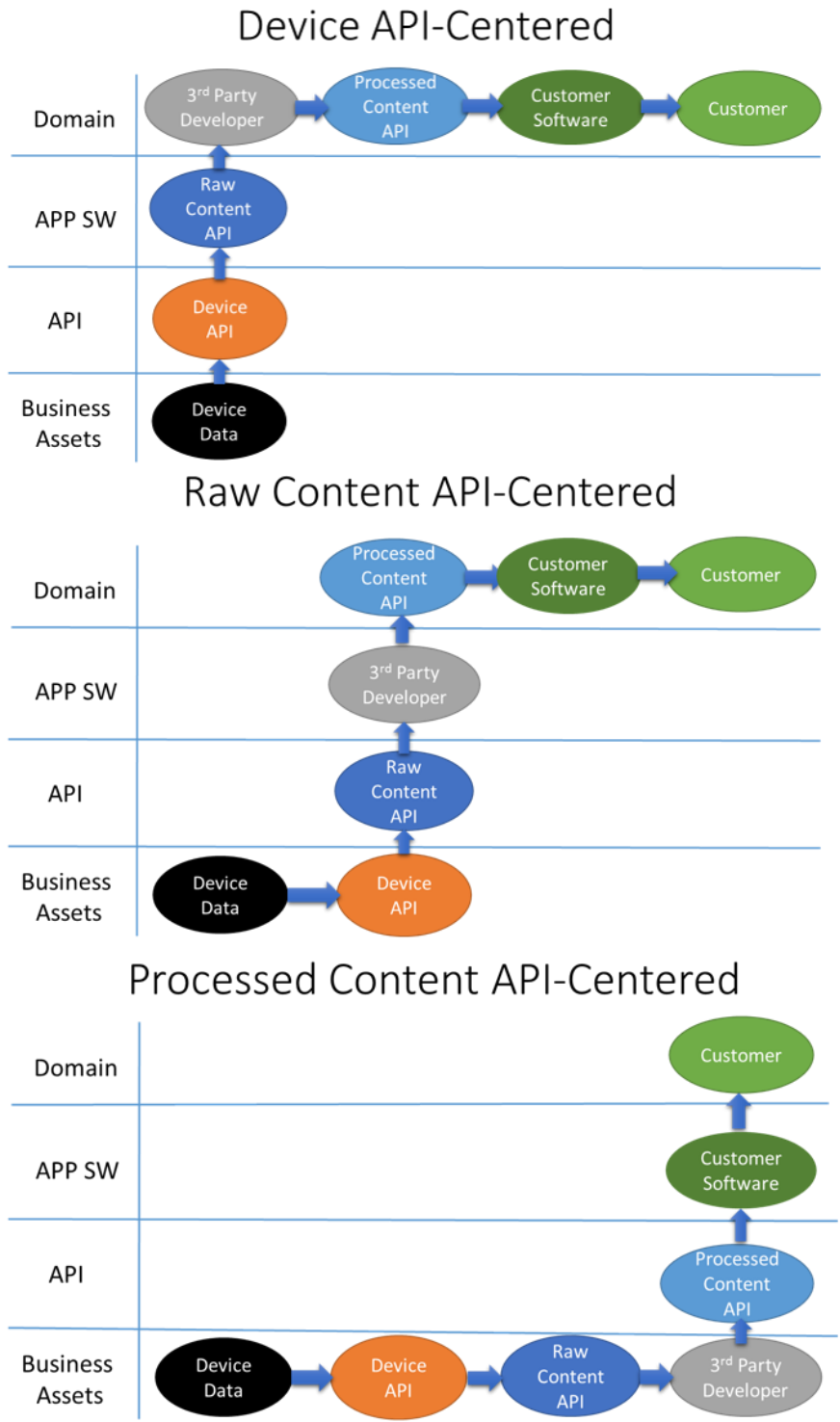

*Figure 21: Simplified view of an Ecosystem Model for Cloud API Mapped to Layers depending on the API of Focus*

### 7.3.5 Further Reading

Horkoff, Jennifer, et al. "Modeling support for strategic API planning and analysis." International Conference of Software Business. Springer, Cham, 2018.

## 7.4 Transforming Value Models to Goal Models

When creating goal models, one can start from scratch as described in Section 7.2, or, if one has created an e3 value model such as in Section 7.1, it is possible to transform the value model to a starting goal model.

### 7.4.1 Intended Benefits
- See section 7.2 for the benefits of goal modeling
- Allows one to create a goal model more easily
- The goal and value models are consistent
- Will find gaps and improve the value model, improving overall API ecosystem understanding

### 7.4.2 When to use this method
When you are interested in making a goal model, but do not know where or how to start, and have already made a value model.

### 7.4.3 When not to use this method
When you are not interested in making a goal model, you already understand the ecosystem well, or when you do not have a value model to transform.

### 7.4.4 Goal Models vs. Value Models

Goal and e3 value models have similarities and differences, and differing strengths.  We summarize these in the table below.  Generally, goal models capture more information than value models, but are more complex.

Table 8:  Comparing Value to Goal Models

| Value Models | Goal Modeling |
| --- | --- |
| Ecosystem Mapping + | Ecosystem Mapping ++ |
| Simpler | More Complex |
| Simpler ontology | Richer ontology |
| Focus on value transitions | Dependencies between actors |
| Focus on quantitative measures | Can capture numbers, can work in qualitative space |
| Analysis with profits | Many types of analysis procedures |
| Can work in "why?" | All about "why?", motivations |
| Not focused on qualities (NFRs) | All about qualities (NFRs) |
| "Or" not well supported | Support for alternatives |

### 7.4.5 Transform Value Models to Goal Models

In the following table, we summarize the mapping from value to goal models.

Table 9: Transforming from Value to Goal Models

| Value Model | Maps to | Goal Model |
| --- | --- | --- |
| Actors | -> | Actors |
| Partnerships | -> | Association links between actors |
| Value activities | -> | Tasks |
| Value flow | -> | Dependencies (+ more?) |
| Value object | -> | Dependums (the thing depending on) |
| Red/orange flows | -> | Unsatisfied goals |
| Stimulus | -> | Goals |

### 7.4.6 How to Apply

Starting with an e3 value model:
1. Draw all value model actors as goal model actors
   a. Goal models do not have actors inside of actors (partnerships), draw all actors separately and add "Part of" association links between them, e.g., actor A is part of actor B.
2. Draw value activities as tasks in the appropriate actor.
3. Draw all value flows as dependencies between the appropriate actors.
   a. The type of dependency (goal, quality, task, or resource) will depend on the type of value flow.
   b. For each value flow, there may also be associated goals/tasks/qualities or resources needed within the actors on either side.
4. If any flows are marked as problematic, they associated dependencies can be marked as "denied" with a black X.
5. Any stimuli in the value model becomes goals in the goal model. They will have to be assigned to one or more appropriate goal model actors.

The result is an incomplete goal model. The modelers should now consider expanding the model in the following ways:
1. Why? Elicit more goals
2. How well? Elicit more qualities
3. How does it all related? Elicit more relationships

Generally, goal model elements should not be "floating" on their own, everything should be attached via dependencies, and/or or contribution links.

### 7.4.7 Examples

The following shows a value model converted to an initial goal model, and then expanded manually.  The first figure shows an overview of the process, which each figure shown in more detail.

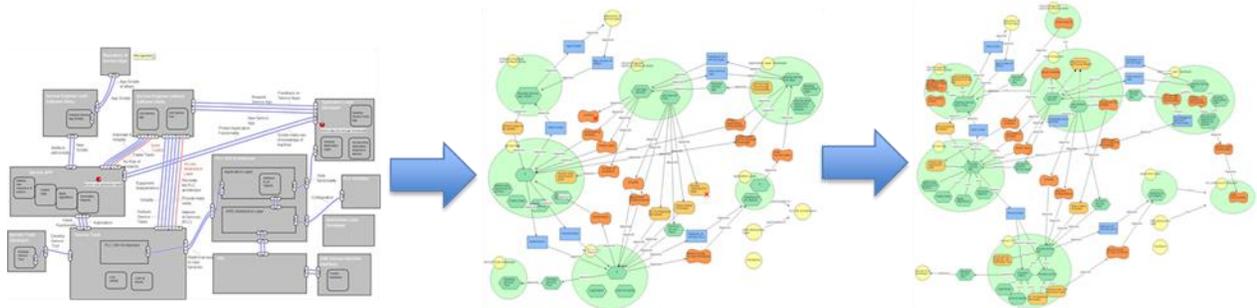

*Figure 22:  Overview of Value to Goal models, value model, initial goal model, then expanded goal model*

# 8   API Operations

## 8.1   API Governance Guidance

### 8.1.1   Benefits

Benefits of this method are
- Identify and evaluate the governance aspects
- Discuss the current and future options for governance models

### 8.1.2   When to use this method

This method is intended for organization who need to plan their API governance process, or who need ideas and feedback to evaluate, improve or verify their current governance processes or structures.

Note:  much of the content of this chapter comes from a thesis, see Section 8.1.9 for references.

### 8.1.3   Overview of Governance Framework

The governance framework is divided into three different components: 1) aspects, 2) strategies and 3) governance board. The work has been carried out in different kinds of organizations that have different levels of governance maturity, so the discussion is relatively general and does not go deep into different roles and governance layers.  Different organizations must adapt and apply the guidance as appropriate.

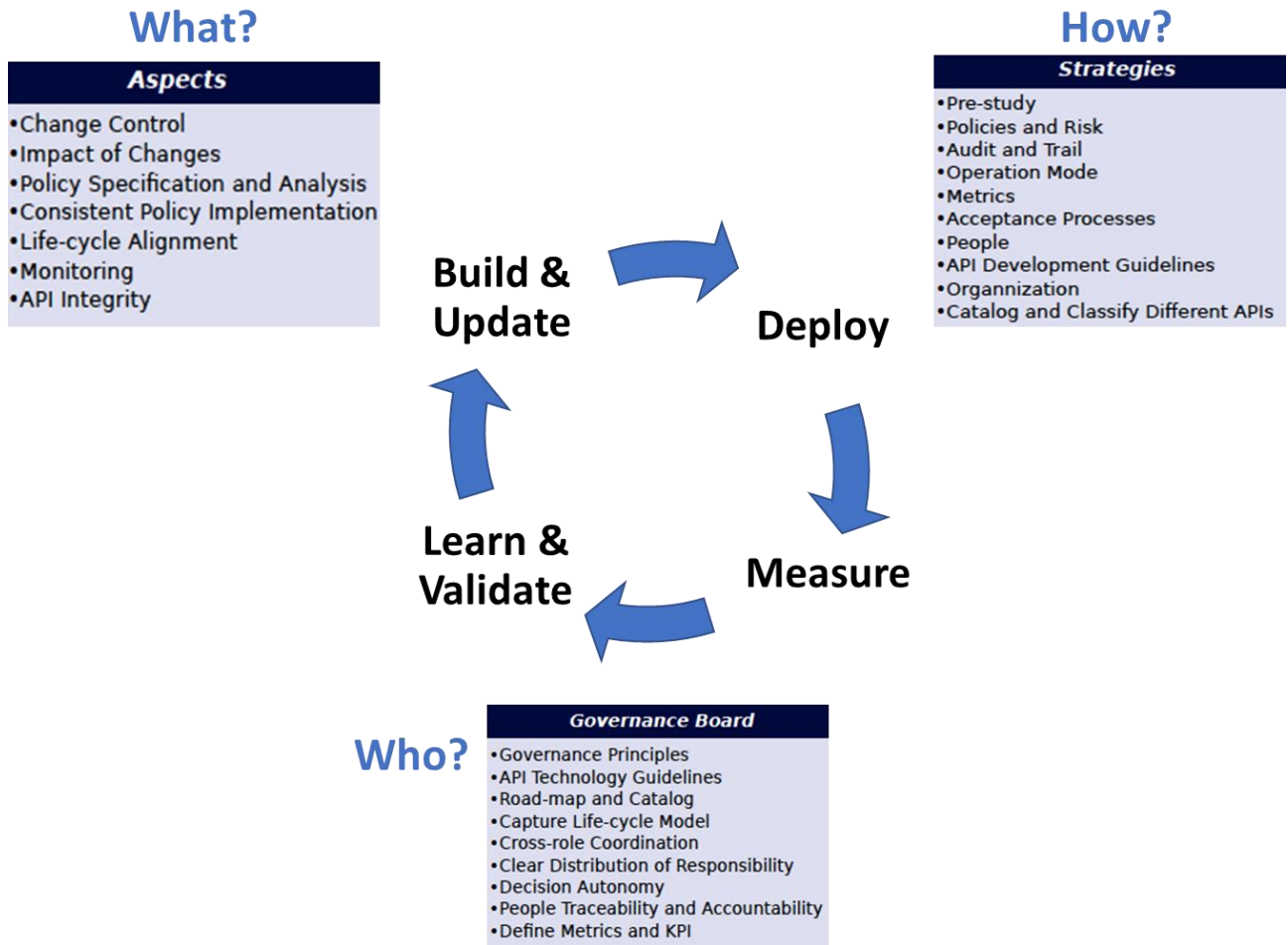

*Figure 7: APIs governance framework*

1) Learn. The status quo. In this phase, the focus is on assessing the current status quo of API governance. From the framework overview, relevant aspects and strategies are identified. By identifying the points that the organization is lacking in, it is possible to move forward to the Implementation phase.

2) Build, Update & Deploy: The new status quo. This phase oversees the implementation of the aspects, strategies and participating roles provided in the overview. Personnel is appointed roles based on key skills. Aspects of governance are considered and analyzed and applied to the governance process.

3) Measure & Learn: Auditing the new status quo. This phase concludes the implementation by producing an audit on how API governance is now performed. Traceability is performed through documentation for future reference and other artifacts. These outcomes are available for re-use in later iterations. When the results are deemed as adequate to the organization's context, the iterative process should then be terminated. However, flexibility of the model allows for a re-evaluation of the status quo and re-start of the activities if necessary.

### 8.1.4 Aspects and Strategies

In what follows, we list aspects and strategies resulting from our findings and which could be considered and deployed by an API governance process.

| Aspects | Description |
|---|---|
| **Change Control** | When changes are required, the resulting effects shall be carefully considered and executed in a consistent method. The long-term vision and purpose of an API should be identified and preserved. The pre-study strategy suggested by this framework helps providing an overview of the expected outcome of a change. Rollbacks to the change should restore functionality also in a consistent and complete manner. |
| **Impact of Changes** | In the context of business, APIs are part of a broad context and any changes to the API can have influence at a business level and IT operations. As such, the impact of changes shall be carefully evaluated. Stakeholders of an API, such as consumers and business owners, shall be informed of changes and what impact they create. |
| **Policy Specification and Analysis** | API governance shall be concerned with access control policies, their analysis and application. Who should access an API, and who should not must be decided taking into account the business context. APIs as access points to assets shall only allow authorized clients access to the asset. |
| **Consistent Policy Implementation** | API governance should ensure that APIs are independent of the technologies that is used to implement the assets. Decoupling API from asset implementation allows for API integrity to be kept: changes to one do not influence the other. |
| **Monitoring** | API governance shall be concerned with monitoring API activity. The proliferation of APIs requires new approaches to control and govern APIs. |
| **Life-cycle Alignment** | The governance process is involved in all the duration of the API life-cycle. For instance, through monitoring API activity, if a decision is made to deprecate the API, the governance process shall ensure that it is not to be awoken again. |
| **API Integrity** | An API shall be able to interface on a newer version of the platform without conflicts, and without effort. When planning new features, existing API should not require extensive refactoring, and backwards compatibility shall be ensured over a period of time. As an interviewee expressed, API governance ensures that "people do not do their own probably incompatible change between other changes". |

*Table 6: APIs governance aspects*

| Strategies | Description |
|---|---|
| **Pre-study** | allows to foresee consequences of requests. A thorough analyses shall be conducted in order to understand the impacts of a change request to the API. The outcome from the analyses is then used to assist the governance board with its final decision. |
| **Policies and risk** | relate to the correct method to work. These strategies are also connected to risk control. Policy strategies ensure that there is compliance with laws, regulations, security control, risk mitigation strategies, corporate guidelines and industry best practices. |
| **Audit and trail** | for instance, via version control systems enables tracking and management of changes to an API. |
| **Operation mode** | establishes a set of goals to guide the development of the API governance process and enable the collection of metrics. For instance, establish what scale of changes the API governance process should be concerned with (small scale changes versus large scale changes). |
| **Metrics** | are the means to provide visibility to governance. The following are suggested metrics to consider: time for changes to be approved, number of approvals and refusals, stakeholder satisfaction scores, technical debt resulting from accepted changes, business impact resulting from accepted changes and service downtime caused by changes. These metrics can be contrasted with the determined policies and operation mode, allowing for an understanding and evaluation of the governance process. |
| **Acceptance processes** | provide the chance to develop and test an API and its compliance with the policies, as well as to make a decision on whether it should be accepted or not. These processes can be automated or human controlled. |
| **People** | should be supported and encouraged to make decisions and requests regarding API changes and features. Who can accept changes, who can request changes, and who can implement changes are all equally important aspects to consider. |
| **API development guidelines** | ensures that APIs are built up to a certain standard. Thoroughly documenting these guidelines ensures that APIs are built in a consistent manner, and can be worked on (developed, maintained) by different people. Additionally, business related roles such as product owners can have a better view of the development process by consulting these documents. |
| **Organization** | is responsible for nurturing a culture of support and reward for good governance over the APIs. |
| **Catalog and classify** | different APIs, providing an accessible method of grouping APIs, for instance with documentation.<br>API versioning could provide a unique identifier to unique states of an API, allowing for clear<br>cataloging and classification. |

*Table 7: APIs governance strategies*

### 8.1.5 API Governance Board

In this section we suggest the creation of an API governance board, with the responsibility to monitor and manage the implementation of the framework and API governance. In the following we list guidelines and input for the formation of such a board.

- **Creation of the governance board.** There are various scenarios in which a governance board could be created. It could be a bottom-up gathering of individuals concerned for development coordination, or it could be a top-down initiative as part of the API creation. In practice, it may be a mix of both, coming from both strategic and practical needs.
- **Composition of the governance board.** Ideally, the board will be comprised of a mix of experience and roles. Most members should have technical experience and familiarity the API and usage SW. Experience in governance or API development is a desired trait. It would be ideal to also have a representative from the business perspective on, or available to the team.
- **Size of the governance board.** The size needed varies depending on the number of developers or users of an API.
- **Governance board principles.** Governance board should be concerned about what principles to follow to ensure the effectiveness of the governance process. Regular assessment of compliance with policies and operation mode strategy defined by the organization should be conducted.
- **API technology guidelines.** Create and maintain guidelines for API technology and development. The governance board should be enabled to make suggestions and decisions on what technology to use in API development, securing API integrity and business value.
- **Road-map and catalog.** Establish a road-map with a well-defined time-frame and goals, and update it regularly, enabling an execution plan with milestones according to the road-map. These could be goals related to pre-study completion, impact of change analysis or establishing an acceptance process for API changes. When building the road-map stakeholder input should be added in order to identify goals that provide business value.
- **Capture life-cycle model.** Create a model that captures the life-cycle of the API, for internal and external API users. If the API is deprecated, the governance board ensures that the API is not awoken again. See Section 6 for a description of how to elicit a lifecycle model.
- **Cross-role coordination.** Ensure that communication channels between the participating roles are kept open to all the interested stakeholders.
- **Clear distribution of responsibility.** Clearly map roles to responsibilities so that everyone in the governance board is fully aware of what they should really do, instead of what they believe they do.
- **Decision autonomy.** Organization should allow the governance board to make important decisions related to API governance. This approach helps avoiding potential bottlenecks by removing the need to constantly report to a single person or "higher ups" from the organization.
- **People traceability and accountability.** Associate people to their respective responsibilities. The concept of "who did what", enables a go-to-person to consult when the understanding of some change or topic is necessary.
- **Define metrics and KPI.** Quantifying business value and metrics increases progress and process visibility to the organization. See Section 8.2-Section 8.5 for guidance on API metrics.
- **Clear rules for escalation**, when necessary.

### 8.1.5.1 Governance Board Examples

Referring to our partner companies, two companies have relatively new APIs and thus are in the process of determining their governance procedures and practices. Three of the companies have a more mature governance process in place. Most of the data gathered on company governance processes was gathered before the current governance framework was created, thus the framework was driven in part by our findings. We provide a brief overview of these cases using text description and excerpts from their value model.

*Technology API*: In this case, API governance is performed by guardians. In this case all system areas, products, areas of functionality have guardians, including APIs. The guardian position is a role which can be played by one or more people in the organization. People can play various roles, including guardians, but often have other roles as well. In this case, they have individual guardians and not teams or boards of guardians. The product and technical guardians meet regularly, discussing topics related to technical debt. There is regular contact between the API guardians and feature owners to discuss changes. API guardians guard both high-level changes and low-level changes using tools like Gerrit. They are often involved in pre-studies, which must be conducted before significant changes. The guardians are helped by design rules, and design documents, which help to capture knowledge on how APIs should be created. The guardian role can be both proactive and reactive.

Part of the need for governance for this company comes as a result of their past agile transition. Previously, teams were component-based, and guarded components by default. Now, as teams are feature base, more guardianship is required, no one "owns" anything. Overall, they are happy with their agile transition, as it improves feedback loops, and gives teams responsibility and authority to deliver more features, but governance and technical debt issues arise as a consequence of the agile setup. Figure 23 shows an excerpt of the e3 value model for this company gives an overview of part of the governance structure, see Section 7.1 for more detail on the value model.

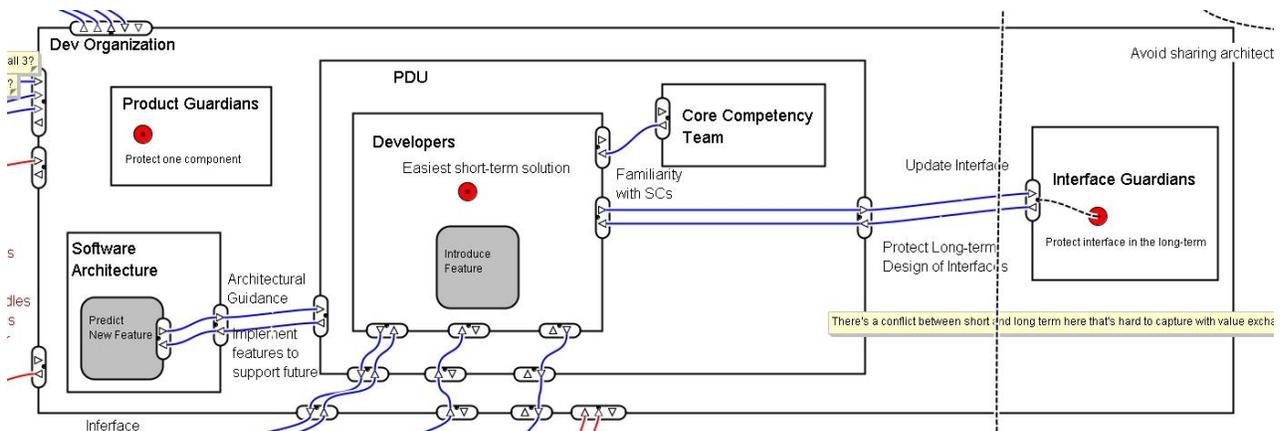

*Figure 23 Excerpt of e3 Value Model for Technology API Showing Governance*

The *Device API* company employs a somewhat similar model in that there are individual API guardians who guard the technical design of the API, including the enforcement of style guides and rules. In this case governance is captured in the value model as an activity to be performed, providing value to the app developers, see Figure 1 for more details, particularly the activity and flow highlighted with a circle.

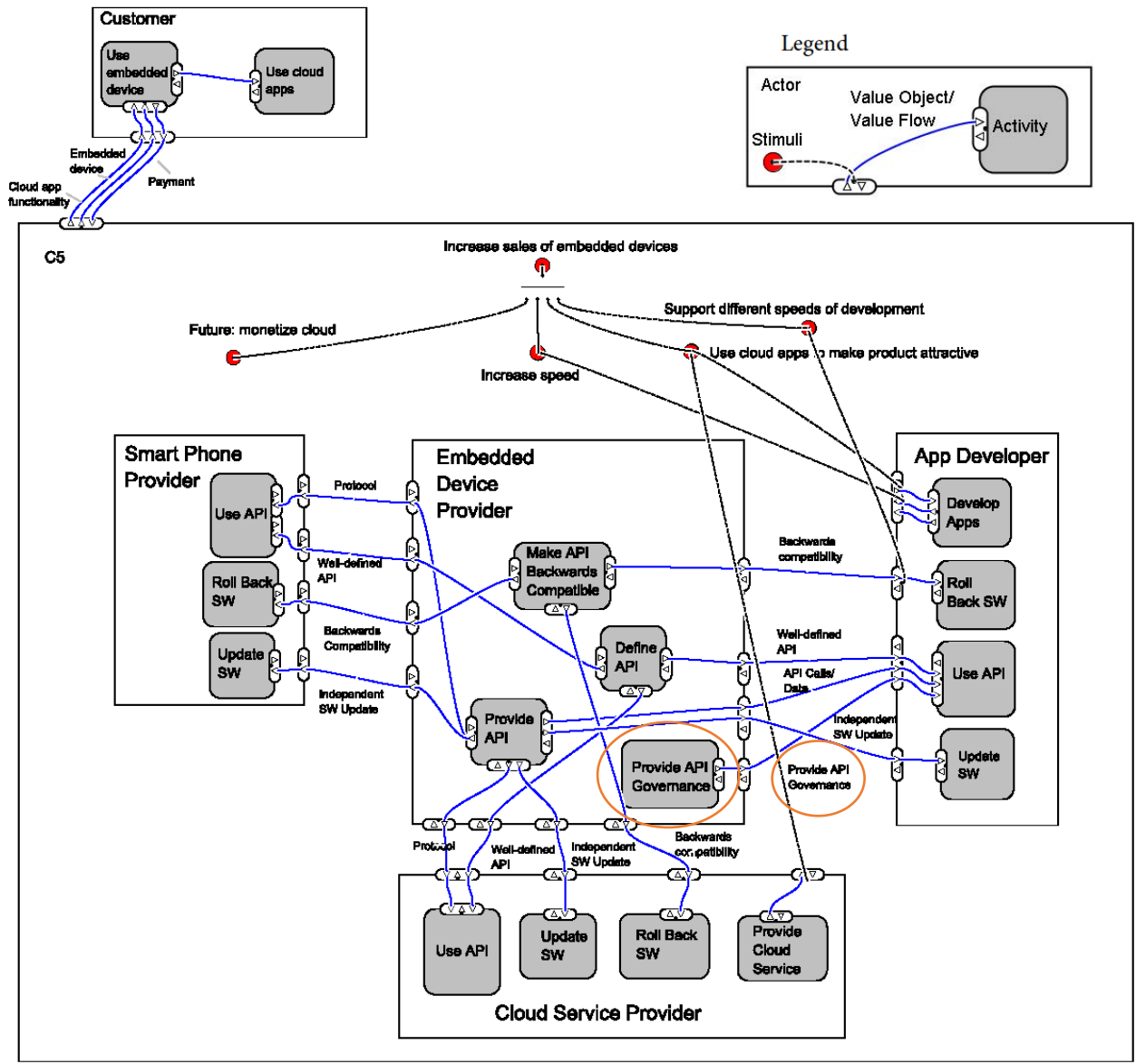

*Figure 24 Excerpt of e3 Value Model for the Device API Showing Governance*

For the *Product API*, API governance takes place as part of the governance board, based on ideas from the concepts of architecture runway from SAFe (https://www.scaledagileframework.com/architectural-runway/).   In this case the board is responsible for monitoring the architecture which supports surrounding rapid agile development.   This is also related to the notion of Agile Release Trains, where the board coordinates agile development.  The governance of the company is represented at a high-level on the bottom right of Figure 25.  In this case the profile (API) owner is in charge of first-line governance of the API, but raises issues and problems regularly with the governance board, which is made up of various technical owners and guardians. The API owner/guardian also represents the area when other issues and problems are discussed, reflecting on the impact to the API.

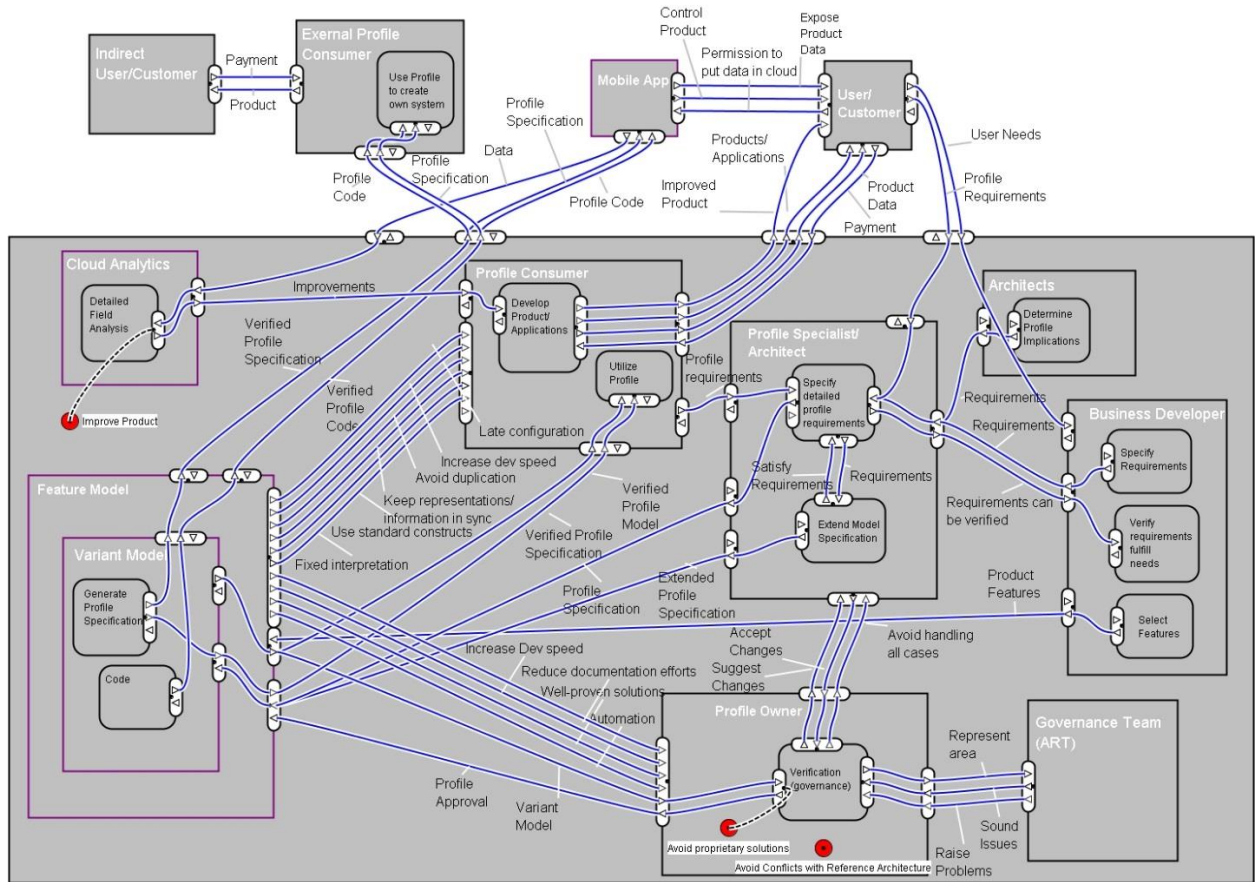

*Figure 25* e3 Value Model for the Product API Showing Governance in the Bottom Right

### 8.1.6   A Model for Determining Governance of API Changes

It is useful to prioritize which APIs, which parts of an API, and which changes should be subject to governance. The scope of change could play a role in this assessment. Bottlenecks can be prevented by ensuring that small scale change requests do not hinder other more impactful change requests from being governed. Furthermore, given the long list of possible governance strategies and aspects to potentially implement, it useful to evaluation and prioritize potential changes or additions to the governance process.   In order to assist in this decision-making process, we suggest models based in the cost-value models, as shown in Figure 26.

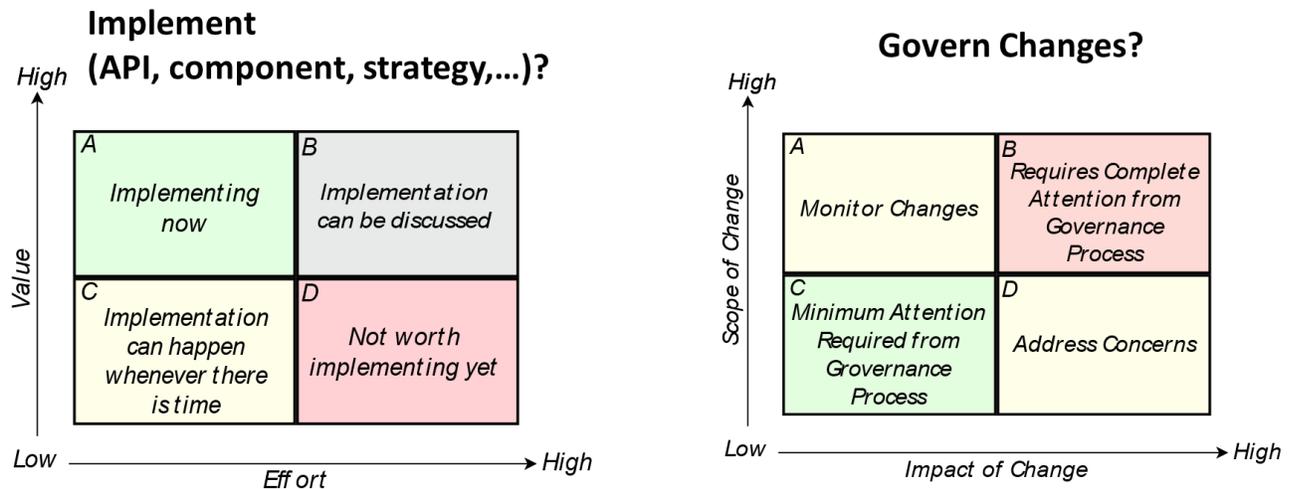
*Figure 7: APIs Governance Decision Model*

In the model, *Value* means the economic or business value associated with the change. *Effort* means costs associated with the change. *Scope of change* means how wide-ranging are the changes planned *Impact of change* means the direct impacts of change for example to customers and other system.  In this case the organization could be the governance board, or the part of the organization forming the governance board.

The left side of the model is the *Implementation Decision Model*.  It helps to identify which aspects and strategies to instantiate, and can also be used to evaluate which APIs or parts of an API to govern. The main concept is to compare the effort that would take to implement a given aspect/strategy versus the value it would bring to the organization. This model allows one to group aspects and strategies to be implemented in a prioritized list.  Likewise, one can compare the effort to govern and API versus the value of the governance, prioritizing where to focus API governance.

The model is divided in sections: A representing what is deemed as most valuable to implement right away, B should be carefully discussed among stakeholders, because it brings high value but takes considerable effort, C refers to non-crucial things that might not be prioritized, and D are red flags that point to aspects and strategies that do not add enough value to the current process considering the effort they take to implement.

The right side of the model is the Change Decision Model.  It helps to identify which changes the API governance process should be concern with. Similar to model presented the left side, this model results in a prioritized list, this time of changes. A comparison is made between the scope of a change against its impact into the system. The result can be grouped in different sections. Section A refers to changes that the governance process should monitor closely, section B should require definitive attention and monitoring, section C are changes that could be ignored as to avoid overwhelming the governance board, and section D are small scope changes to be discussed only if monitoring is necessary or attention is required.

These decision models are intended to be easy to use and help the process of deciding what to govern. Decisions are encouraged to be taken by gathering all the API stakeholders together. Once together, discussions can be initiated such as on how to prioritize the presented aspects and strategies according to the organization. In what follows is a summary of a decision-model.

### 8.1.7 How to use this method

The suggestions and material in this chapter can be applied by the governance board, if one exists, or by those who would form or create such a board, depending on the maturity of the organizations' API governance process.

**API Openness:** consider the users of the API, the level of openness and exactly who and what should be governed as part of the governance process. These considerations relate both to Section X, the layered architecture of the API and Section Y, ecosystem mapping. Understanding the level of openness of the API can help to select and form government aspects and strategies.

**Aspects and strategies:** consider the aspects and strategies from Section 8.1.6. Which aspects or strategies resonate as relevant or important? Discuss how these strategies could be implemented at the organization. If there are too many relevant aspects or strategies to consider, use the change model to prioritize the potential changes or additions to API governance.

**API Governance board:** consider the recommendations for the governance board. Determine which are relevant and useful for the organization. As above, if too many recommendations are useful to make at once, prioritize using the change model.

**Change Model:** If there are too many relevant aspects and strategies or governance board recommendations, use the left side of this model to prioritize changes. This side of the model can also be used to prioritize APIs or parts of APIs to govern, if the API landscape is complex.
Consider potential changes to potential parts of the API, and use the model to determine if they should be governed by the governance board.

Questions that may help in recognizing the scope and impact:
- How many internal or external customers the change will affect?
- Is the API critical?
- Is the API strategic?

### 8.1.8 Governance Discussion

API governance is challenging topic, and although this chapter gives some guidance, there is a lot which remains unaddressed. Specifically, we cover several issues raised by our conversations with our industrial partners.

**Agile Methods.** There is an inherent conflict between agile methods and governance. Most agile methods (e.g., scrum) advocate for team autonomy, while the governance of APIs may conflict with this autonomy. There must be a continual balance made between agility and the coordination facilitated by **APIs.** This balance will be affected by the particular development process used by the organization, e.g., scaled agile frameworks like SAFe.

**Terminology.** "Governance" is often a hard term to sell, particular in light of the perceived conflict with agility. It implies a top-down set of restrictions which can be perceived negatively. The organization may want to explore different terminologies facilitating different mindsets, e.g., facilitation, coordination, enabling, or educating.

**Lifecycle.** There is a strong link between governance and the lifecycle model in Section 6. The lifecycle characteristics table covers this to some extent, indicating that the level of governance, and the form of governance will change depending on the lifecycle of the API. Generally, initially governance will be rather focused and light, with the more normal complete form of governance during operation, with governance activities changing and tapering off during depreciation and retirement.

**Governance vs. Planning.** A distinction can be made between governing an existing API and planning the creation or redevelopment of an API. These activities are linked, planning an API should involve planning aspects of API governance. Although the focus of this chapter is in governance, elements of API planning and development are also included. Sometime such planning activities are performed outside of the boundaries of API governance, i.e., they are dictated by management or marketing. An organization will have to adapt the governance framework when applying it to their particular situation, including omitting planning elements which are out of their control.

**Automation.** Thus far, this section has not discussed whether governance tasks or rules can or should be automated. Particularly when the API or number of developers is large, automation can be key to effective governance. Ideally, rules for API changes can be encoded and checked automatically, as part of or similar to various software testing practices. Automatic checks of rules could produce warnings or errors depending on the severity of the issue. A mechanism for dealing with rule deviations must be created, and this mechanism could be a mix of manual or automatic processes (e.g., sending notifications, blocking check-ins or integrations, in person discussions). Although automation is desirable, realistically, many rules or desirable API properties will not be able to be encoded in an automatic fashion, so the process will have to be designed to support a mix of manual and automatic actions. For more on automated quality checks of APIs, see the later section on on automating design metrics, Section 8.5.

### 8.1.9 Further reading

1. Emanuel Lourenço Marcos, Raphael Puccinelli de Oliveira, A Framework for Guidance of API Governance: A Design Science Approach, Bachelor of Science Thesis in Software Engineering and Management, 2018
2. Ostrom, Elinor, et al. *Rules, games, and common-pool resources*. University of Michigan Press, 1994.
3. Leffingwell, Dean. *SAFe® 4.0 Reference Guide: Scaled Agile Framework® for Lean Software and Systems Engineering*. Addison-Wesley Professional, 2016.
4. R. Wolski, C. Krintz, H. Jayathilaka, S. Dimopoulos, and A. Pucher, "Developing Systems for API Governance," Sept 2013. [Online]. Available: https://figshare.com/articles/Developing Systems for API Governance/790746
5. H. Jayathilaka, C. Krintz, and R. Wolski, "EAGER: Deployment-Time API Governance for Modern PaaS Clouds," in 2015 IEEE International Conference on Cloud Engineering, March 2015, pp. 275–278.
6. C. Krintz and R. Wolski, "Unified API Governance in the New API Economy," Cutter Consortium, 37 Broadway, Suite 1 Arlington, MA 02474, U.S.A., Tech. Rep., Sept 2013
7. C. Krintz, H. Jayathilaka, S. Dimopoulos, A. Pucher, R. Wolski, and T. Bultan, "Cloud Platform Support for API Governance," in 2014 IEEE International Conference on Cloud Engineering, March 2014, pp. 615–618.

## 8.2 Determining API Metrics

### 8.2.1 Intended Benefits
- Examples of API metrics used in practice
- Finding potential API aspects to measure
- Understanding the challenges specific to API measurements

### 8.2.2 When to use this method
When one is interested in measuring qualities and aspects of an API at various levels and doesn't necessarily know what to measure, or wants further ideas.  Start with this method if you have not produced a goal model as in Section 7.2.  If you have a goal model, consider skipping to Section 8.3.

### 8.2.3 When not to use this method
When API measurement is too expensive to implement, uninteresting, or already established.

### 8.2.4 Description

Measuring the operations of an API from a strategic and ecosystem perspective can provide insight into the health of the API.   We have made progress in identifying different classes of API measures as per our layered architectures, and have identified metrics which are typically more easy or difficult to measure.

#### 8.2.4.1 Example Metrics

Table 10 lists metrics extracted from the brainstorming sessions from the companies.  The rows here do not have a particular meaning. We group these metrics into related categories in the text below the table.

We can make several observations about the collected metrics.  First, many are not metrics per se, they do not have an immediate quantification, e.g., product impact on brand.  Others, although not directly quantifiable have a list of associated metrics from existing software metric standards, e.g., reliability.  Others are more easily quantifiable in the derived context, e.g., num. revisions.  The metrics are at very different levels of abstraction, which makes them easier or harder to measure.   In some cases they are desired qualities which could be measured.

| Company 1 | Company 2 | Company 3 | Company 4 | Company 5 |
|---|---|---|---|---|
| Service use data | Num using tool directly | Version control vs. not | Num times features used | Short vs. long-term costs |
| Compliance to privacy laws | Num use cases covered | Frequency of updates | Technical debt metrics | Versioning |
| Product impact on brand | Num functions used | Num. Revisions | Cost saved by reusing features | Risk, Security |
| Num. regulations satisfied | Log changes to levels | Failed commits | Amount of functions in products | Compliance |

| % system with old vs. new firmware | | Failed tests in backwards compatibility | Integration problems before/after | Reliability |
|---|---|---|---|---|
| Time taken to upgrade | | Complexity metrics | Usability and understandability | Stability |
| Num. customer opt-ins | | API LOC change vs. new features | Business risk of failure | Extendibility |

*Table 10: Sample Metrics Categories Collected During Workshops*

**Sample Metrics Grouped by Category**

**Data Usage**
Service use data

**Security, Privacy, Law**
Compliance to privacy laws
Num. regulations satisfied
Risk, Security
Compliance

**Branding**
Product impact on brand
Business risk of failure

**Deployment**
% system with old vs. new firmware
Time taken to upgrade
Num. customer opt-ins
Integration problems before/after release

**Usage**
Num using tool directly
Num functions used

**Functionality**
Num use cases covered
API LOC change vs. new features
Amount of functions in products

**Evolution**
Change logs, frequency of change
Frequency of updates
Num. Revisions, Versioning
Failed commits
Failed tests in backwards compatibility

**Costs**
Short vs. long-term costs
Cost saved by reusing features

**API Qualities**
Complexity metrics
Technical debt metrics
Usability and understandability
Reliability
Stability
Extendibility

### 8.2.5  How to Apply

First, one can try a "bottom-up" method. Using ideas from the section above, think of relevant metrics for your API of choice?
- How can these be measured?
- Are they being measured currently?
- Is this knowledge useful for your API?

Review Table 10 to see if any of these metrics may apply in your case.

When metrics ideas are exhausted, try a "top-down" method.  If you already have a goal model, turn to Section 8.3 and follow the instructions to link metrics to goals.  If you do not have a goal model, try to list the main goals of the API ecosystem.  For each goal, ask "how can we measure this goal?"  If the goal is too abstract to measure, ask "how can we break this goal down?" until something more measurable is reached.

Map to Roles
For each metric, ask "who?".   Who is interested in this metric and its result?

Map to Sources
For each metric, if possible, as "from where?"   What is the data source or information source(s) for this metric?

Suggestion:   Put the above collected information into a table, such as the following.

| Metric (What?) | Goal (Why?) | Roles (Who?) | Sources (from where?) |
|---|---|---|---|
|  |  |  |  |

### 8.2.6   Summary
Measuring various aspects of an API can provide critical strategic information.  Although API metrics relate to metrics for general software and ecosystems, they are unique and should be considered.

### 8.2.7   Further Reading

Lagerstedt, Robert. "Using automated tests for communicating and verifying non-functional requirements." *Requirements Engineering and Testing (RET), 2014 IEEE 1st International Workshop on*. IEEE, 2014.

Implementing automatic tests for architectural rules" by Johan Bäckström and Fredrik Karåker Sundström, Masters Thesis, Lund, 2017

## 8.3   Linking Metrics to Ecosystem Models

### 8.3.1   Intended Benefits
- Understanding why metrics are being measured, how they match to strategic goals
- Understand who receives metric information and why

### 8.3.2   When to use this method
When one is interested in evaluating the utility of metrics, verifying they are used by someone for some purpose.  To use organization goals to derived metrics.   Also understanding how metrics fit into and aid in an API ecosystem.   Use this method when you have produced a goal model from Section 8.3, or if you have a list of API goals.

### 8.3.3   When not to use this method
When the reasons, use and role of API metrics are clear.  If goals are unknown.

### 8.3.4 Metrics Hierarchy

Thus far, we have treated the qualities one may wish to measure over an API as relatively independent. Realistically, these qualities related to each other in various ways. To consider these relations, one can draw a hierarchy of qualities related to APIs, qualities which can be potentially measured. This is similar to the general hierarchies of non-functional requirements for systems provided historically by Boehm or by the NFR Framework (see references in Section 8.3.8).

Figure 26 shows an example for one of our partners with the Device API. This uses some of the goal model syntax from Section 7.2 (qualities, contribution links). The qualities start rather abstract and broad at the top of the hierarchy and become more concrete and measurable near the bottom. Here, the overall quality to achieve is profit. The left side focuses on qualities from the perspective of the API user, while the right side focuses on the qualities of the API developer. The user wants to use the API to support revenue, but providing customer value and customer effort. Customer effort can be decomposed into maintainability, learnability, usability, and readiness, and further decomposed into design stability and documentation. Developer effort/cost is decomposed into implementation stability, security, and legal compliance, further composed into more detailed measures. The bottom layer of qualities are relatively consistent with the metrics provided in Table 12 in Section 8.5.

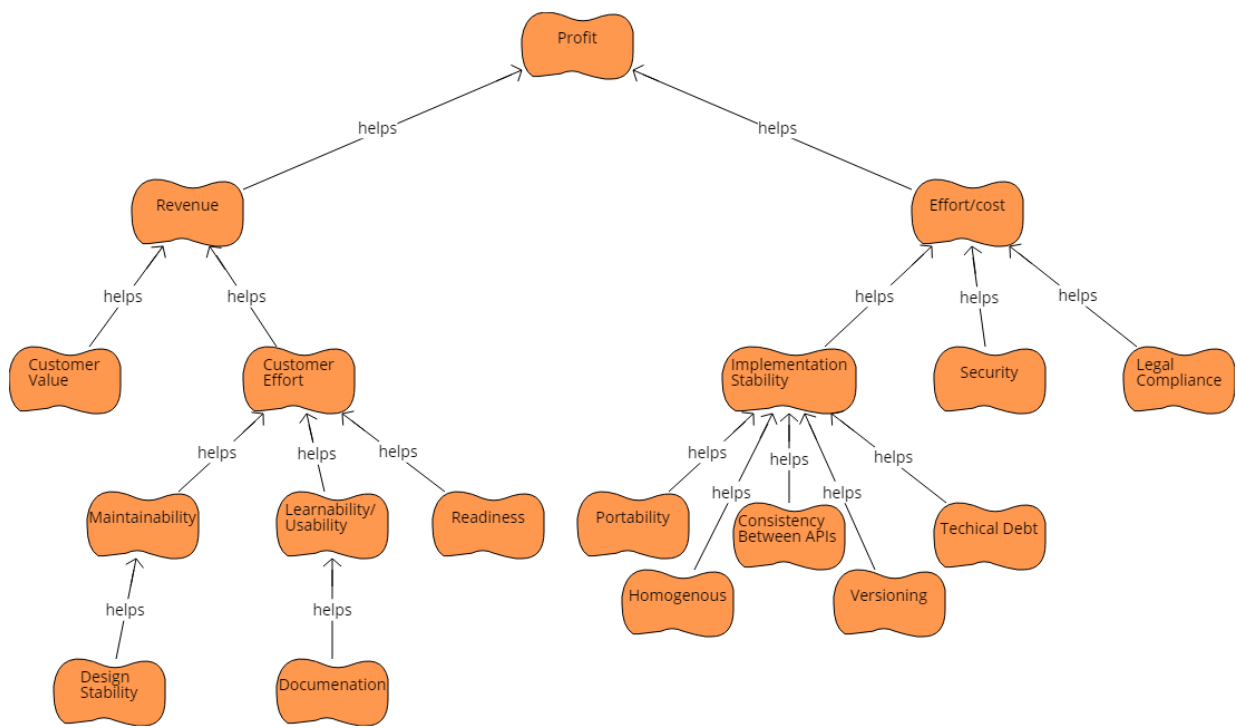

*Figure 26: Example of API Quality Hierarchy from the Device API*

Each company's set of API qualities/metrics and the resulting hierarchies may differ.

### 8.3.5 Mapping to Ecosystem Models with Goal Models

In order to determine how an organization's API metrics fit into an API ecosystem, and to understand why metrics are collected, we can map collected API metrics and qualities to a company's ecosystem map. If an organization has gone through Section 7.2, they will have a resulting goal model capturing the

API ecosystem. If this method was skipped, one can still map metrics to goals by coming up with individual goals or tasks for each metric. This process is similar to what is advocated in the Business Intelligence Modeling framework, (see references in Section 8.3.8 and Figure 27).

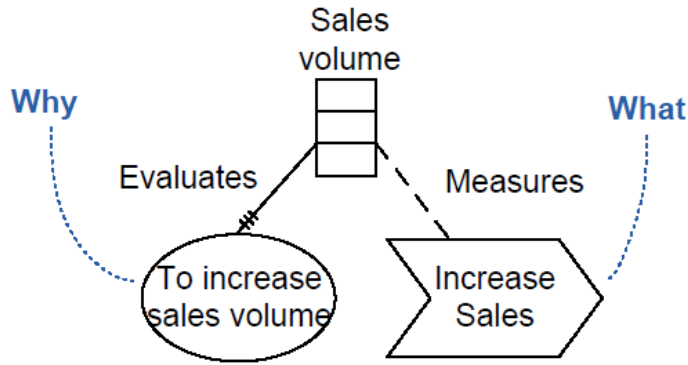

Figure 27: Figure from Horkoff et al. 2014 Showing a Metric (Sales Volume) linked to a Goal and a Task

An example goal model for a company is shown in Figure 28. This captures the actors, goals, qualities, tasks and dependencies in an API ecosystem. One can map each quality/metric found in the previous sections to goals and tasks in the ecosystem model, answering "why?" or "how?". The overall process is shown in Figure 29, with a simplified version of the result shown in Figure 30.

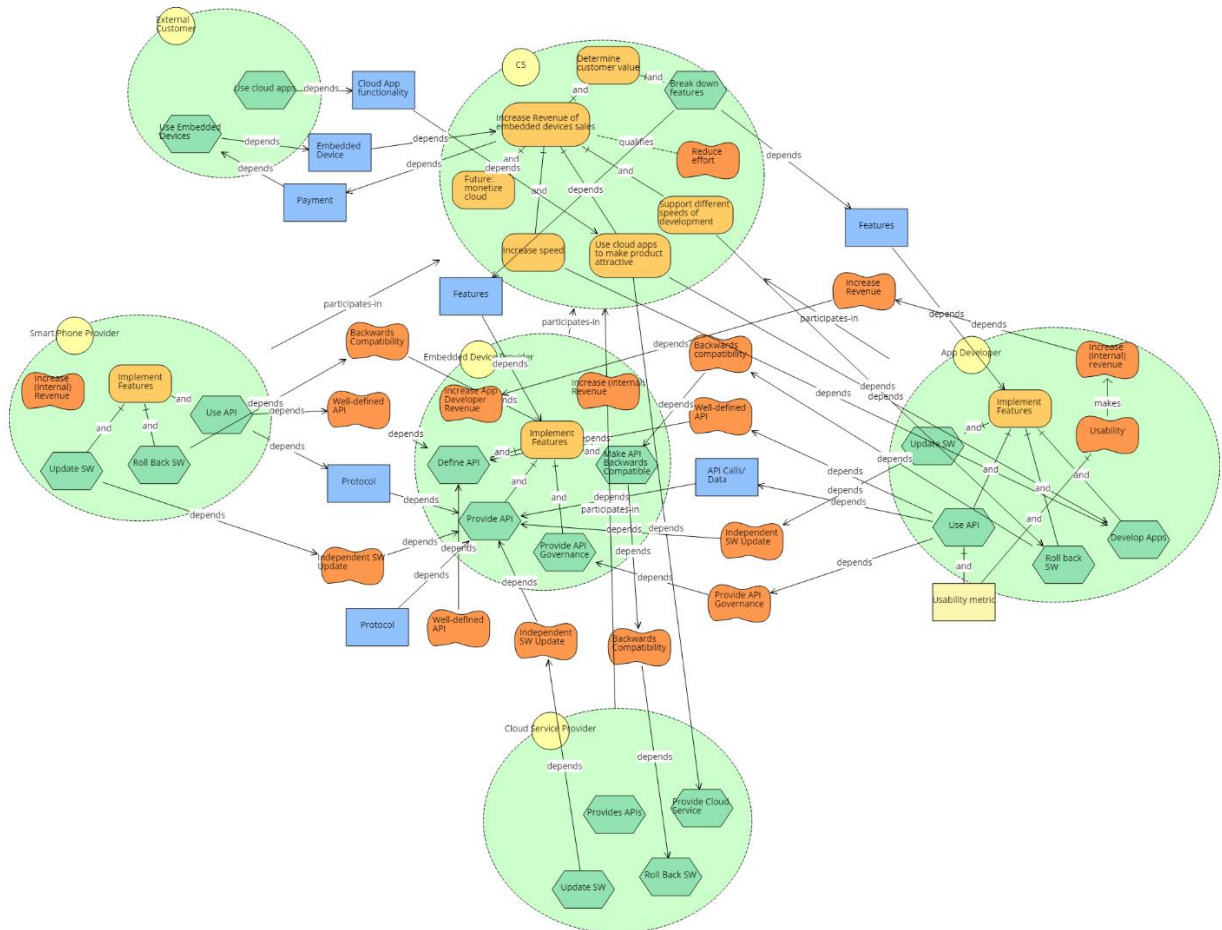

Figure 28 An Example Ecosystem Model for the Device API

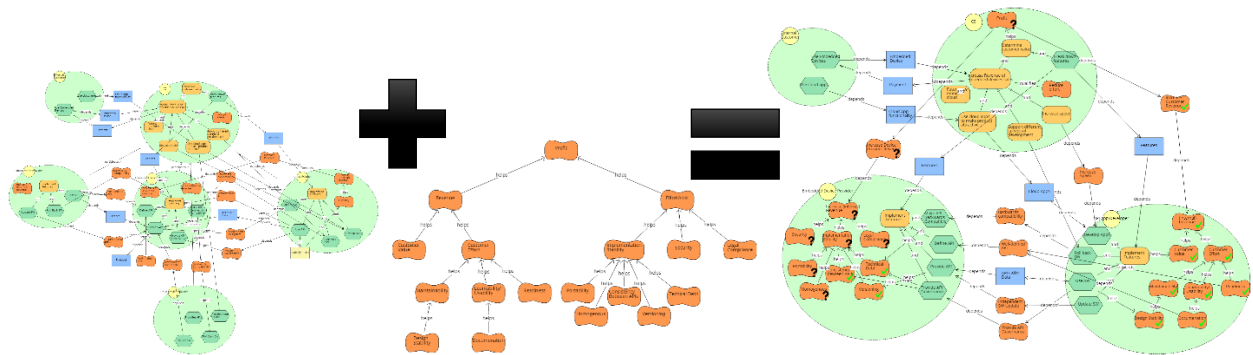

*Figure 29 The Process of Mapping Metrics/Qualities to the Ecosystem Goal Model*

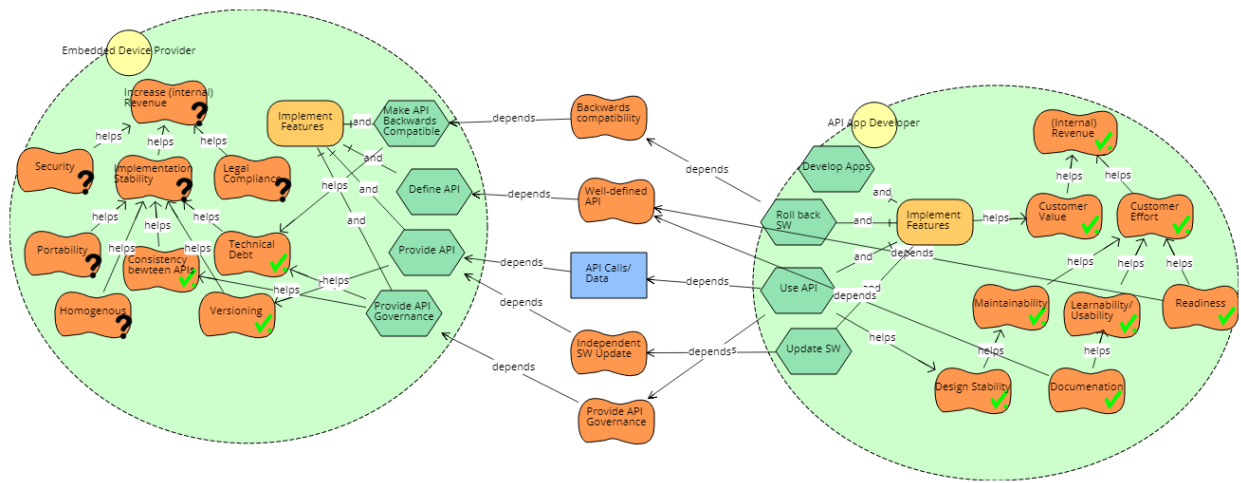

*Figure 30 An Example Ecosystem Model from E with Metrics/Qualites Linked to Goals/Tasks*

The final ecosystem model includes an evaluation of which qualities/metrics are satisfied or not are satisfied via their link to goals/tasks in the model. For example, the task Provide API Governance helps ensure Consistency between APIs, so this quality is marked with partially satisfied (check dot). These values are propagated up the metric hierarchy when possible, sometimes resulting in an unknown value (?) when it is not clear how some qualities are satisfied.

The mapping between API qualities and company tasks and goals, as well of the assessment of which qualities are achieved, will differ between companies.

### 8.3.6   How to Apply

There are two major steps: mapping out a metric hierarchy and mapping the hierarchy to goals and tasks, possibly from an ecosystem model. The first step is optional, and for the second step, the use of an existing ecosystem model is also optional.

Step 1:   Consider the metrics you have deemed relevant in previous exercises.   Can they be placed in a hierarchy?
- What is the top-level metrics/qualities that drive success?

- What metrics contribute to other metrics?  The contributors can be at a lower level of the hierarchy?
- What are the metrics that can be directly measured?  These are at the bottom of the hierarchy?
- Are there any metrics in between the bottom and the top?

Step 2:  For each metric above, ask "Why?" and "how?" questions.
- Why is the metric measured, what goal is it trying to achieve?
- Is there an associated task or action that allows one to collect the measurement?
- These goals and tasks can be elicited from scratch or taken from an existing ecosystem model.
- If new goals and tasks are discovered, it can be helpful to add them to the ecosystem model, if it exists, in order to make the other model more complete and accurate.
- Can the goals and tasks be linked to other metrics/qualities, or other goals and tasks?  Do these goals and tasks "belong" to a particular actor?  Answering these questions can help to develop a draft ecosystem model including metrics.
- Overall, the role of the metrics in the ecosystem should become clearer.

### 8.3.7 Summary
It is useful to understand why and how for API metrics, including how these metrics contribute to the big picture.

### 8.3.8 Further reading

Boehm, Barry, and Hoh In. "Identifying quality-requirement conflicts." IEEE software 2 (1996): 25-35.

Chung, Lawrence, et al. Non-functional requirements in software engineering. Vol. 5. Springer Science & Business Media, 2012.

Horkoff, Jennifer, et al. "Strategic business modeling: representation and reasoning." Software & Systems Modeling 13.3 (2014): 1015-1041.

## 8.4 Layered Metrics

### 8.4.1 Intended Benefits
- Classifying API metrics into the different API layers for better understanding of metrics
- Helping to find potentially missing metrics
- Understanding the impact of metrics
- Provide a checklist of potential API metrics at each level

### 8.4.2 When to use this method
When one is interested in finding further metrics or understanding the impact of metrics on various levels.

### 8.4.3 When not to use this method
When API measurement is too expensive to implement, uninteresting, or already established.

### 8.4.4 Metrics Categories

In a workshop discussion with a case company particularly interested in metrics, it was suggested by the company API experts that the various API metrics could be classified as per the API dimensions as shown in Figure 31.

The **API Business** dimension can contain metrics which measure API-related business aspects, such as revenue, increase market share, increase in customers, and other strategic objectives.  This could include further metrics, such as compliance to privacy laws, impact on brand, and short vs. long-term costs.

The **API Usage** dimension measures aspects concerning the software that uses the API.  These can fall into the regular categories of software metrics, including modularity metrics, stability, frequency of changes, bug rates, etc.  This dimension can contain more general usage statistics such as number of users, or number of users of particular API calls.  Other measures like number of functions used, and amount of functions in products could be assigned to the Usage Statistics dimension.  Forwards and backwards compatibility of the API could be measured at this level.

The **API Design** dimension relates to the design of API itself, including metrics such as number of API call parameters (related to complexity), division of API calls into sub-section (related to modularity).  Other examples include number of revisions, failed commits, and qualities such as reliability and stability.

The **API Implementation** dimension covers the applications which implement the API, measuring aspects related to instance implementation such as conformance with API conventions.  As the API implementation is software, regular software metrics can apply.  Forwards and backwards compatibility of the API could also be measured at this level.

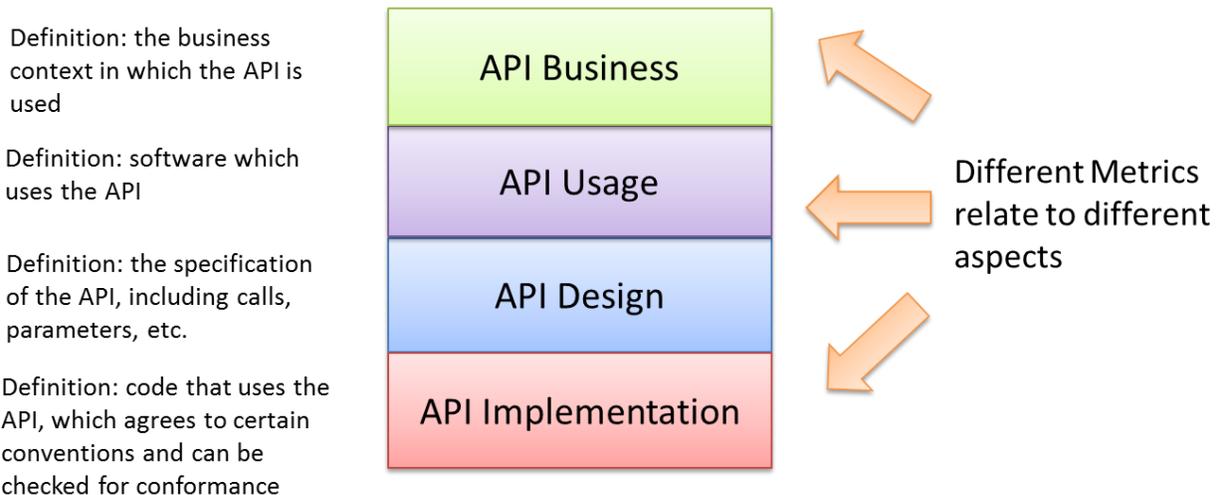

*Figure 31  API Metric Layers and their Definitions*

Questions to help to elicit metrics in these dimensions are show in Figure 32.

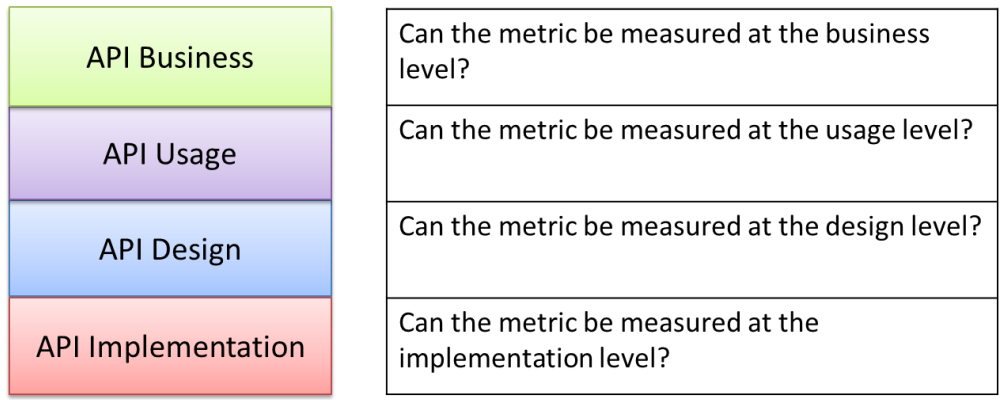

Figure 32 API Metric Layers with Associated Questions

In Table 11, we show sample high-level metrics and whether or not they could possibly map to each of the four API metric Dimensions. This mapping was gathered as per a discussion with one company, thus the mapping could differ between companies.

| Metric | API Business | API Usage | API Design | API Implementation |
|---|---|---|---|---|
| Short vs. long term costs | Split (too general) | | | |
| Maintenance/ Maintainability | | | Yes | Yes (different metric) |
| Documentation | | | Yes (relevant) | |
| Decommissioning (to determine) | Yes (revenue) | Yes (usage) | | |
| Versioning (# released versions) | | Yes | Yes | Yes |
| Compliance (legal) | | Yes | Yes | Yes |
| Dev speed (sub metrics, usable) | | Yes | | Yes |
| Quality | | Yes | Yes | Yes |
| Reliability (failures, data accuracy) | | Yes | | Yes |
| Stability (frequency of change) | Yes (diff) | Yes (diff) | Yes | Yes (different) |
| Branding | Yes | | | |
| Learnable | | Yes | Yes | |
| Usable | | Yes | Yes | |
| Interestingness | Yes | Yes | | |
| Homogenous | Yes | yes | Yes | |
| Satisfaction | Yes | Yes | | |
| Performance | | Yes | | Yes |
| Implementation Portability | | Yes | Yes | |
| Readiness – when releasable | | Yes | Yes | Yes |
| Extendibility | | Yes | Yes | Yes |
| Technical debt | | Yes | Yes | Yes |

| Compatibility (versions compatible) |  | Yes |  |  |
| Security |  |  | Yes | Yes |
| Risk | Yes | Yes |  |  |

*Table 11: Sample Metrics for each API Metric Dimension*

### 8.4.5 How to Apply

Use case 1:  Identify further metrics
Find relevant/interesting metrics for each dimension in Table 11.   Use the questions in Figure 32 to help the discovery process. As with above:
- How can these be measured?
- Are they being measured currently?
- Is this knowledge useful for your API?

Use case 2:  Identify the impact of metrics
It may be difficult to find API-related metrics for each layer.  Instead, one could make a list of which metrics influence or effect each layer, for example, the results of a metric measured on the API implementation  may have an effect on the business layer, or not.   Go through the metrics gathered thus far and consider their impact.
- At which layer are the metrics measured?
- Which other layers do they impact?

### 8.4.6 Summary
APIs can be measured at both the technical and strategic level, and at various conceptual dimensions.

## 8.5 Automating Design Metrics

### 8.5.1 Intended Benefits
- Considering metrics specific to the design layer
- Consideration of potential level of automation of API metrics

### 8.5.2 When to use this method
When one is interested in design-level API metrics, particularly in automating these metrics.

### 8.5.3 When not to use this method
When measuring API design is not important or challenging, or automation is not desired.

### 8.5.4 API Design Metrics

Although it can be important to find and measure metrics at every dimension, metrics at the API design level are often the most challenging to measure, as unlike implementation or usages software, the API signature design cannot be executed.  Thus the measurements are over static constructs.  However, these measures can often be quite important, as one wants to understand qualities of the API design before release, after which modifying the design is difficult.

In this section we consider which metrics can be applied at the API Design dimension, and particularly whether or not these metrics can be automated, or even partially automated, or whether they must be applied manually. We describe this in Table 12. Here the 2nd column considers whether the metric can be automated, partially automated, or must be conducted manually. Notes to justify this consideration, potentially with examples, are provided. This information comes from one company, so as with Table 11, could differ across companies.

For example, in row one, we consider whether one could measure the quality "maintainability" over the API design layer. In this case, there are ways to measure extendibility, e.g., avoiding default values in method signatures. Such measurements can be automated, but other qualities of maintainability, e.g., modularity, comprehensibility, may be more difficult to measure.

| Metric | Automate at API Design | Example/Notes |
|---|---|---|
| Maintenance/ Maintainability | Partially Automatable | Extendibility, can you add without breaking backwards compatibility. A few rules, like avoiding default values in a method. |
| Documentation | Automatable | Completeness - all in and out parameters documented. |
| Versioning (# released versions) | Automatable | Number of versions, number of APIs, number of variants, number released. |
| Compliance (legal) | Partially Automatable | Detect if it's providing email or personal numbers, some of this can be automated. |
| Stability (frequency of change) | Automatable | How often does the design change? |
| Learnable | Manual | Can do manually. Learnable: do we understand what this method is doing? Need to evaluate with people. |
| Usable | Manual | Usable: how would I as a developer use this? Difficult to test automatically, need user testing. |
| Homogenous | Partially Automatable | E.g., API with 2 values: put value, set b. Within one API. Partially automatable with guidelines. |
| Implementation Portability | Partially Automatable | Use for different implementation. e.g., get service status, must not be measurement dependent (don't expose implementation specific data). |
| Readiness – when releasable | Partially Automatable | Beta testing + stability. |
| Technical debt | Partially Automatable | Guidelines, rules and recommendations, breaking a rule introduces technical debt. Some rules are mandatory (legal rules), many are not. |
| Security | Partially Automatable | Manually: All APIs have permission tag, making sure they have the right tags, e.g., personal data tags. In a guideline, checks manually. E.g., password in clear text. |
| Consistency (between APIs) | Partially Automatable | Style consistency between APIs. Most on single APIs, this on collection of APIs. |

*Table 12: Consideration of Metrics at the API Design dimension, including Automation Possibilities*

### 8.5.5 How to Apply

Consider metrics specifically at the API design level, not having access to the API usage software or implementation.  Is it important in your case to check API design quality before release? As with above:
- How can these be measured?
- Are they being measured currently?
- Is this knowledge useful for your API?

Consider particularly:
- Can these metrics be automated?  Semi-automated?  Or must they be performed manually?

### 8.5.6 Summary

Metrics at the API design level are particularly difficult to automate, but can help to ensure quality API design before deployment.

# Appendix

## 8.6 Methodology

In this section, we briefly describe the methods used to develop the framework, empirical research conducted closely with our partner companies.  More details can be found in our published papers.

The research employed multiple case study methodology. The case data were collected from workshops, interviews, and thematic discussions with experts from four companies in the embedded systems industry. We worked with the companies to develop frameworks for their immediate needs regarding the management of APIs. We first collected perceptions from the key stakeholders and then combined these insights with knowledge gained via literature reviews, providing and receiving feedback to and from our partners concerning their specific challenges over several workshops.

In what follows, we list aspects and strategies resulting from our findings and which could be considered and deployed by an API governance process.

The project has evolved over four sprints (2 years).  Generally, in a sprint we conduct the following activites to develop framework topics and methods:

> **Initial cross-company workshop:** This two to three hour workshop involves one to two representatives from each company. Here we solidify sprint plans.
>
> **Optional Preparatory Meetings:** In some cases we have preparatory meetings with companies, typically remotely, particularly in situations where we are working with a new company, or where the example of focus is new.
>
> **Material Preparation:** Between workshops, the researchers use their knowledge of the cases to apply the various topics and methods and create draft material to use in workshops, to be validated by the companies.

**Company Workshops:** We attended an in-person workshop at the location of each company. The workshops are three hours long, except in the cases where a preparatory workshop had taken place, then they are two hours. Workshops are attended by anywhere from one to six company representatives, with an average of about three to four attendees. Attendees are selected by the main company representatives, usually those who have detailed knowledge of the API and can represent various roles (users, designers, guardians). During the workshops the researchers and company representatives work together to develop, apply and evaluate the various methods of the APIS framework

**Post-Workshop Iterations:** After the individual workshops, we update the material to reflect received feedback, then sent the resulting material back to the participants via email. In some cases we receive further feedback and make further iterations, while in other cases the companies have no further feedback.

**Mid-Sprint Workshop:** Halfway through the sprint we conduct a one to two hour workshop to report intermediate progress and discuss early findings. Sprint plans are adjusted if necessary.

**Final Cross-Company Workshop:** To wrap-up the sprint, we have a final three-hour cross-company workshop where we presented all developed material and methods for all the companies, and discuss findings, comparing results. The company representatives had a final opportunity to correct the sprint outcomes. Plans for the next sprint are discussed.

**Thesis Supervision.** In some sprints the project involves the participation of Bachelors or Masters students conducing their thesis on a topic related to the project.

**Reporting.** Project findings and framework topics and methods are reported through project slides, reports such as this one, and academic publications.